\documentclass[12pt]{article}
\newif\ifams\amsfalse                                                    
\amstrue                                                                 
\ifams                                                                   
 \usepackage{amsfonts}                                                   
\fi                                                                      
\newif\iffigs\figsfalse                                                  
\figstrue                                                                
\newif\ifdraft\draftfalse                                                
\ifdraft
 \else
\fi
\title{}

\newif\ifinter\interfalse                                               

\ifdraft
  \def\secl#1{\nopagebreak\marginpar{\vspace{-6mm}\scriptsize #1}\label{#1}}
  \def\beql#1{\marginpar{\vspace{4mm}\scriptsize #1}
              \nopagebreak\begin{equation}\label{#1}}
  \def\ftl#1#2{\footnote{\label{#1}[#1] #2}}
  \def\bibl#1{\marginpar{\vspace{4mm}\scriptsize #1}\nopagebreak\bibitem{#1}}
 \else
  \def\secl#1{\label{#1}}
  \def\beql#1{\begin{equation}\label{#1}}
  \def\ftl#1#2{\footnote{\label{#1}#2}}
  \def\bibl#1{\bibitem{#1}}
\fi

\def\draftnote#1%
  {\ifdraft
    \hspace{1mm}\newline\noindent\begin{tabular}[t]{|p{14cm}|}
     \hline {\bf DRAFT NOTE}\\ #1 \\ \hline
    \end{tabular}\newline\noindent
   \fi}

\def\supnote#1%
  {\ifdraft
    \hspace{1mm}\newline\noindent\begin{tabular}[t]{|p{14cm}|}
     \hline {\bf Supplementary Note}\\ #1 \\ \hline
    \end{tabular}\newline\noindent
   \fi}

\def\internote#1%
  {\ifdraft\ifinter
    \hspace{1mm}\newline\noindent\begin{tabular}[t]{|p{14cm}|}
     \hline {\bf Supplementary Internal Note}\\ #1 \\ \hline
    \end{tabular}\newline\noindent
   \fi\fi}

\def\multdn{$\downarrow\downarrow\downarrow\downarrow\downarrow$}
\def\beginsup%
 {\noindent\begin{tabular}[t]{|c|}
   \hline {\bf Supplements}\\ 
   \multdn\multdn\multdn\multdn\multdn\multdn
   \multdn\multdn\multdn\multdn\multdn\multdn \\ \hline
  \end{tabular}}

\def\multup{$\uparrow\uparrow\uparrow\uparrow\uparrow$}
\def\endsup%
 {\noindent\begin{tabular}[t]{|c|}
   \hline \multup\multup\multup\multup\multup\multup
          \multup\multup\multup\multup\multup\multup \\
   {\bf Supplements}\\ \hline
  \end{tabular}} 

\iffigs
 \input epsf
 \def\pct#1{\centerline{ \epsfbox{#1.ps}}}
\else
 \def\pct#1{(see figure in file #1.ps)}
\fi
\newif\ifappend\appendfalse
\appendtrue
\ifappend
 \newcommand{\newsection}[1]{
  \vspace{10mm} \pagebreak[3]
  \refstepcounter{section}
  \setcounter{equation}{0}
  \message{(\thesection. #1)}
  \addcontentsline{toc}{section}{\protect\numberline{\arabic{section}}{#1}}
  \begin{flushleft}
   {\large\bf \thesection. #1}
  \end{flushleft}
  \nopagebreak}

\else
 \newcommand{\newsection}[1]{\section{#1}}
\fi

\def\al{\alpha}

                \def\Dl{\Delta}
\def\ep{\epsilon}
\def\kp{\kappa}
\def\lm{\lambda}               \def\Lm{\Lambda}

\def\th{\theta}               
\def\vph{\varphi}

               \def\Sg{\Sigma}

\def\Sc{\mbox{\protect$\cal S$}}

\ifams
 \def\bbl#1{{\mathbb #1}}
\else
 \def\bbl#1l{{\bf #1}}
\fi
\def\CC{\bbl{C}}

\def\RR{\bbl{R}}
\def\ZZ{\bbl{Z}}

\def\diag{{\rm diag}}
\def\im{{\rm Im}}

\def\min{{\rm min}}

\def\tr{{\rm tr}}

\def\pt{\partial}
\def\goto{\rightarrow}

\def\dg{^{\dagger}}
\def\inv{^{-1}}

\def\rec#1{{\raise 0.4ex \hbox{$\scriptstyle {\frac{1}{#1}}$}}}

\def\derp#1{\frac{\pt}{\pt#1}}
\def\half{{\raise 0.4ex \hbox{$\scriptstyle {1 \over 2}$}}}
\def\vev#1{\left\langle#1\right\rangle}

\def\hs{\hspace{5mm}}
\def\hsm{\hspace{2mm}}
\def\hsc{\hspace{5mm},\hspace{5mm}}
\def\ie{{\em i.e.}}

\def\beq{\begin{equation}}
\def\eeq{\end{equation}}

\def\Wtree{W_{\rm tree}}
\def\Lml{\Lm_L^{b_L}}
\def\Lmr{\Lm_R^{b_R}}
\def\Lmd{\Lm_D^{b_D}}
\def\Lmlp{{\Lm'_L}^{b'_L}}
\def\Lmrp{{\Lm'_R}^{b'_R}}
\def\sce{\stackrel{\rm SC}{=}}
\def\NS{N\hspace{-0.8mm}S_5}
\def\nl{\newline}
\def\nlb{\newline $\bullet$ }
\def\st{{\rm st}}

\def\NPB#1{Nucl. Phys. {\bf B#1}}
\def\PLB#1{Phys. Lett. {\bf B#1}}
\def\PRD#1{Phys. Rev. {\bf D#1}}
\def\PRL#1{Phys. Rev. Lett. {\bf #1}}
\begin{document}


\begin{titlepage}

\ifdraft \fbox{\bf !!!!!!!!!!!!!! DRAFT VERSION !!!!!!!!!!!!!!}\vspace{-1cm}
\fi

\begin{flushright}
RI-8-97\\
hep-th/9708168\\[5mm]
\ifdraft
 \count255=\time 
 \divide\count255 by 60
 \xdef\hourmin{\number\count255}
 \multiply\count255 by-60
 \advance\count255 by\time
 \xdef\hourmin{\hourmin:\ifnum\count255<10 0\fi\the\count255}
 \number\day/\number\month/\number\year\ \ \hourmin
\\[15mm]\fi
\end{flushright}

\begin{center}
\Large
{\bf\boldmath M Theory, Type IIA String\\
  and 4D $N=1$ SUSY \\ 
  $SU(N_L)\otimes SU(N_R)$ Gauge Theory}
\\[10mm]
\large
Amit Giveon \normalsize and \large Oskar Pelc
\normalsize
\\[5mm]
{\em Racah Institute of Physics, The Hebrew University\\
  Jerusalem, 91904, Israel}\\
E-mail: giveon@vms.huji.ac.il, oskar@shum.cc.huji.ac.il
\\[15mm]
\end{center}
\begin{abstract}
$SU(N_L)\otimes SU(N_R)$ gauge theories are investigated as effective field
theories on $D_4$ branes in type IIA string theory. The classical
gauge configuration is shown to match quantitatively with a corresponding
classical $U(N_L)\otimes U(N_R)$ gauge theory. Quantum effects freeze the
$U(1)$ gauge factors and turn some parameters into moduli.
The $SU(N_L)\otimes SU(N_R)$ quantum model is realized in M
theory. Starting with an $N=2$ configuration (parallel $NS$ fivebranes), the
rotation of a single $NS$ fivebrane is considered. Generically this leads to a
complete lifting of the Coulomb moduli space. The implications of
this result to field theory and the dynamics of branes are discussed. When the
initial $M$ fivebrane is reducible, part of the Coulomb branch may survive.
Some such situations are considered, leading to curves describing the 
effective gauge couplings for $N=1$ models. The generalization to models with
more gauge group factors is also discussed.
\end{abstract}
\draftnote{
BEFORE SUBMITION:
\nl Print: Preprint Nr.; Acknowledgments; Spell Check
\nl Source: heading; draft flag; $\{\}$From; line overflow}

\end{titlepage}

\ifdraft
 \pagestyle{myheadings}
 \markright{\fbox{\bf !!!!!!!!!!!!!! DRAFT VERSION !!!!!!!!!!!!!!}}
\fi

\flushbottom


\newsection{Introduction}

The realization of supersymmetric gauge field theories as theories 
describing the low energy dynamics of branes in string theory 
\cite{Polchinski}, is an approach 
that led to much progress in the understanding of both subjects. 
To study $N=4$ SUSY gauge theories in 3 dimension,
a particularly
simple and useful construction was introduced in \cite{HW9611}, involving
type IIB $D$ branes and $\NS$ branes in flat spacetime. 
To consider $N=1$ SUSY gauge theories in 4 dimension, a 
construction
involving type IIA $D$ branes and $\NS$ branes in flat spacetime was 
presented and studied in \cite{EGK9702}.
These constructions were
generalized in several directions \cite{BHOOY9612}-\cite{AOT9708}, leading to
realizations of field theories
in various dimensions and with various amount of unbroken supersymmetry.
In particular, one may consider type IIA $D_4$ branes with boundaries on $\NS$
and/or $D_6$ branes \cite{EGK9702}. 
The field theory describing the low energy dynamics of the 
$D_4$ branes is a 4D gauge theory%
\footnote{It is 4D rather then 5D, because the $D_4$ branes are finite in one
  of their spatial dimensions, so this dimension is invisible in the long
  wavelength approximation.}.
$N_c$ parallel $D_4$ branes suspended (in one of their directions)
between two parallel $\NS$ branes will lead to $N=2$ SYM $U(N_c)$ theory;
adding
$D_6$ branes, or semi infinite $D_4$ branes, corresponds to fundamentals in the
field theory, leading to $N=2$ SQCD; more than
two $\NS$ branes lead to a gauge group with more $U(N)$ factors.
A rotation of some of the branes breaks the supersymmetry further, and an 
appropriate rotation leads to $N=1$ supersymmetric models. Such rotations
correspond in field theory to turning on masses for some of the adjoint
chiral superfields (chiral components in the $N=2$ vector multiplets).

The above description was obtained in the limit of small string coupling
constant, (where string perturbation theory is reliable). The branes are
considered to be flat and one ignores interactions between them.
One also considers energy scales below the string scale, so that all the
massive modes of the string can be ignored.
For a sufficiently small string coupling, the gauge coupling
constant is also small and, consequently, one obtains correspondence with a
{\em classical} field theory. Quantum corrections to the classical field 
theory correspond to quantum corrections in the dynamics of the branes.
In particular, it was shown in \cite{Witten9703} that a $D_4$ brane ending on
a $\NS$ brane causes the latter to bend. One of the implications to the
corresponding field theory is that the $U(1)$ factor in the $U(N)$ gauge
group is frozen.
Another effect was identified through a comparison with known quantum results
in field theory. It was shown in
\cite{EGKRS9704} that when the $\NS$ branes are not parallel 
there are quantum forces between $D_4$ branes, 
corresponding in field theory to a dynamically
generated superpotential.

The weakly coupled type IIA string is a particular limit of M theory.
A systematic study of many aspects of the quantum 
behavior in a limit of M theory corresponding
to a strong string coupling
was initiated in \cite{Witten9703};  
the string coupling turns into a large 10'th spatial direction, and 
the $D_4$ and $\NS$ branes configuration of the type IIA limit 
turns into a worldvolume of the M theory fivebrane $M_5$.
This approach was followed also in
\cite{Kol9705}-\cite{AOT9708}.
One considers a small Planck length in 11
dimensions (relative to the characteristic scales of the configuration),
and this justifies the use of a low energy (long wavelength) approximation.
The fivebrane worldvolume
is of the form $\RR^{3+1}\times\Sg$, where $\Sg$ is a Riemann
surface, and it can be determined by imposing appropriate asymptotic
conditions, which represent the characterization of the quantum system
considered.
Once the curve is known, it provides some information.
The space of solutions (corresponding to given asymptotic conditions), 
represents the moduli space of vacua of the quantum system, so one obtains in
this way the number of vacua (or the dimension of the moduli space, when it is
continuous). Note that this should be  the {\em quantum} moduli space, possibly
modified by a dynamically generated superpotential. This means that the 
M theory description is expected to take into account the effect that in type
IIA string theory was interpreted as a force between $D_4$ branes (as mentioned
above). The genus of $\Sg$ is the number of massless Abelian gauge
fields in the low energy effective field theory and when it is non vanishing,
$\Sg$ determines the effective gauge coupling -- it is precisely the
Seiberg-Witten (SW) curve \cite{SW} of the corresponding gauge theory.
The curves for $N=2$ models with a simple
classical gauge group and fundamental matter were found 
\cite{Witten9703,LLL9705,BSTY9705}, in complete agreement with the known
results \cite{HO9505}-\cite{AS9509}. This method was also used to obtain
unknown results. In particular,
curves for several $N=2$ SUSY models with product gauge groups were
derived \cite{Witten9703,LLL9705,BSTY9705,LL9708}.

In this work we continue the study of the realization of 4D SUSY
gauge theories in the type IIA string and M theory.
Some of our goals are to fill gaps in the ``dictionary'' relating string and
field theoretical phenomena and to use this dictionary to obtain information
about both subjects.
We consider models based on $U(N_L)\times U(N_R)$ and $SU(N_L)\times SU(N_R)$
gauge groups, with bi-fundamentals and possibly also with fundamentals and/or
adjoints. Such models are realized
by three $\NS$ branes connected by $D_4$ branes (and additional semi-infinite
$D_4$ branes). We concentrate on branches of vacua with vanishing vev's of
the fundamentals. Other aspects of this model were considered in \cite{BH9704}.
In section \ref{FT}, we perform a field-theoretical analysis.
We start with the classical analysis of both models, emphasizing the
differences between them. We find that the main difference is that the
Fayet-Iliopoulos parameters in the $U(N)$ models transform to moduli in the
$SU(N)$ model. We also compute the resulting moduli spaces of vacua.
Next, we consider quantum corrections to the moduli space, in situations that
are investigated later in M theory.
In section \ref{IIA},
we consider the brane configuration in type IIA string theory.
First the ``classical'' description is given.
It should correspond to the classical $U(N_L)\times U(N_R)$ model and we
indeed display a detailed, quantitative, geometric identification of almost
all the parameters and moduli of the field theory. We list the evidence for
this identification. In particular, a perfect match is
found in the moduli spaces. The only parameters that seem to have no
geometrical manifestation are the Yukawa couplings of the bi-fundamentals.
We then describe the modifications
in this picture caused by quantum effects. In particular, the $U(1)$ factors
are frozen, which means that the low energy effective theory has an
$SU(N_L)\times SU(N_R)$ gauge group.

In section \ref{M-th},
we move to the strong string coupling limit -- M theory --
looking for the $M_5$ brane that corresponds to the type IIA brane
configuration. We start with an $N=2$ supersymmetric configuration
-- parallel $\NS$ branes. 
In this configuration there is a Coulomb branch, parameterized
by the vev's of the adjoint fields.
The general form of the corresponding SW curve was 
determined in \cite{Witten9703}. We find the explicit curve, \ie, its 
dependence on the coordinates and moduli, by considering various limits.
We then consider what happens when one of the branes is rotated (breaking the
supersymmetry to $N=1$). We find that
generically (\ie, when the curve is irreducible), the full Coulomb branch is
lifted. This remains so for models with more $\NS$ branes (more gauge group
factors). In field theory this means, for example, that when a mass is given
to the adjoint field of the first gauge factor, all the moduli are fixed,
including those corresponding to the other adjoints. 
This differs from the classical situation,
where only that part of the moduli space corresponding to the massive
adjoint is lifted. One of the implications of this result is that a Coulomb
branch can survive only when the curve is reducible, which means that the
polynomial that defines it is factorizable. We explore this possibility and
then investigate in detail one such situation in which the central $\NS$
brane, detached from the rest of the configuration, is rotated. This leads to
SW curves for $N=1$ models with $SU(N)\otimes SU(N)$ gauge group with
fundamentals and bi-fundamentals (and no adjoints)%
\footnote{Curves for $N=1$ models are known only for a limited number of cases:
  \cite{IS},\cite{EFGR},\cite{HO9505},\cite{EFGIR}-\cite{Gremm9707}.}.
Some of these curves
were derived before by field-theoretical considerations \cite{CEFS,Gremm9707}
and we find complete agreement with them. For other models our results seem to
be new.
Finally, we return to the irreducible case and analyze in detail the
rotation of the left $\NS$ brane in the absence of fundamentals, obtaining
explicit solutions for the curve. We check the limit of vanishing gauge
coupling for the right factor and recover correctly the results for SQCD found
in \cite {HOO9706}. We conclude in section \ref{disc}  with a discussion.
In an appendix, we explain our strategy in the use of symmetries.

\newsection{The Field Theoretical Models}
\secl{FT}

In this section we present and analyze the field-theoretical models that
emerge from the brane configurations discussed in the later sections.

\subsection{The Classical \boldmath $U(N_L)\times U(N_R)$ Model}

We consider an $N=1$ supersymmetric gauge theory with a gauge group 
$U(N_L)\times U(N_R)$, two bi-fundamentals and for each factor of the gauge
group, an adjoint and several fundamentals ($n_L$ and $n_R$ respectively).
The matter content is described in the following table:
\beq\begin{array}{|l||c|c||c|c|c||c|c|c|}
 \hline \rule[-1.3ex]{0em}{4ex}
        &F^a_{\hsm\bar{a}} &\tilde{F}^{\bar{a}}_{\hsm a}
        &{A_L}^a_{\hsm b} &{Q_L}^a_{\hsm i} &\tilde{Q}^{\hsm i}_{L\hsm a}
        &{A_R}^{\bar{a}}_{\hsm\bar{b}} &{Q_R}^{\bar{a}}_{\hsm\bar{i}} 
        &\tilde{Q}^{\hsm\bar{i}}_{R\hsm\bar{a}} \\ 
 \hline \rule[-1.3ex]{0em}{4ex}
 U(N_L) &[N_L]_+ &[\bar{N}_L]_- 
        &[Adj]_0\oplus[1]_0 &[N_L]_+ &[\bar{N}_L]_- 
        &[1]_0 &[1]_0 &[1]_0 \\ 
 \hline \rule[-1.3ex]{0em}{4ex}
 U(N_R) &[\bar{N}_R]_- &[N_R]_+ &[1]_0 &[1]_0 &[1]_0  
        &[Adj]_0\oplus[1]_0 &[N_R]_+ &[\bar{N}_R]_- \\
 \hline\end{array}\eeq
Here $a=1\ldots N_L$ and $\bar{a}=1\ldots N_R$ are color indices,
$i=1\ldots n_L$ and $\bar{i}=1\ldots n_R$ are ``flavor'' indices; 
$[R]_q$ denotes a representation of $U(N)=SU(N)\otimes U(1)$: a representation
$[R]$ under $SU(N)$ and a $U(1)$ charge $q$. The superpotential considered is 
\begin{eqnarray}\nonumber
  W & = & \xi_L\tr A_L+\half\mu_L\tr A_L^2+\tr(m_L\tilde{Q}_LQ_L)
          +\tr(\lm_L\tilde{Q}_LA_LQ_L)+ \\ \label{Wtree}
    &   & +\xi_R\tr A_R+\half\mu_R\tr A_R^2+\tr(m_R\tilde{Q}_RQ_R)
          +\tr(\lm_R\tilde{Q}_RA_RQ_R)+ \\ \nonumber
    &   & +m_F\tr(\tilde{F}F)+\kp_L\tr(\tilde{F}A_LF)
          +\kp_R\tr(FA_R\tilde{F})
\end{eqnarray}
(in matrix notation)%
\ftl{Redund}{In some situations, some of the parameters are redundant. For
  example, when $\mu\neq0$, one can eliminate $\xi$ by a (scalar) shift of 
  $A$, while when $\mu=0$, such a shift can be used to eliminate $m_F$. This
  will be also apparent in the corresponding brane configuration, to be
  described later.}.
We also allow Fayet-Iliopoulos (FI) terms for the $U(1)$
vector superfields $\tr V_L$, $\tr V_R$
\[ \int d^2\th d^2\bar{\th}(\eta_L\tr V_L+\eta_R\tr V_R) \hsc
   \eta_L,\eta_R\in\RR \hs. \]
\internote{$N=2$ and $SU(2)_R$: 
  \nlb breaking conditions; 
  \nlb identify $U(1)_J$;
  \nlb $\xi,\eta$ as an $SU(2)_R$ triplet (in $N=2$ SUSY description).}
\internote{Is it consistent to treat the adjoint and singlet of $A$ together?
  Is this respected by quantum corrections? In the present work this question
  is irrelevant, since after quantization we actually consider a model 
  without the singlets.}

We will determine the (classical) vacua of this model,
restricting our attention to vacua with%
\footnote{For generic values of $m_L,m_R$, this is actually implied by the 
  equations of motion.}
$Q=0$. 
The $D$-term equations for a supersymmetric vacuum are 
\beql{D-eq}
  [A_L,A_L\dg]+FF\dg-\tilde{F}\dg\tilde{F} = -\eta_LI_{N_L} \hsc
  [A_R,A_R\dg]+\tilde{F}\tilde{F}\dg-F\dg F =-\eta_RI_{N_R} \hs.
\eeq
(where $I_N$ is the $N$ dimensional identity matrix) and they imply%
\footnote{The vanishing of $[A,A\dg]$ is explained in \cite{APS9603}; the rest
  of the above results can be obtained following the procedure described in
  the appendix of \cite{CEFS} for a similar model.}:
\begin{itemize}
\item $[A_L,A_L\dg]=[A_R,A_R\dg]=0$;
\supnote{PROOF \cite{APS9603}:
  \nlb In squaring the D term, one can show that the cross terms cancel.
  \nlb $[A,A\dg]$ is an $SU(2)_R$ singlet, while the rest is a
  part of a triplet (together with $dW/dA$ and its conjugate).
  \nl IMPLICATIONS:
  \nl $A$ can be diagonalized by a color rotation, which is essential for the
  brane interpretation. Note, however, that in the following we use the gauge
  freedom to diagonalize $F,\tilde{F}$ and not $A$. The F-term equations will
  imply that in all cases $A$ will be also diagonalizable, simultaneously with
  $F,\tilde{F}$.}
\item $\eta_L=-\eta_R=:\eta$;
\item if $N_L\neq N_R$ then $\eta$ must vanish;
\item $\eta$ can always be chosen non-negative
  (possibly after a $L\leftrightarrow R$ transformation);
\item $F$ and $\tilde{F}$ can be simultaneously diagonalized by a color 
  rotation and then they assume the following form
  \beql{fix} F=\diag\{c_a\} \hsc \tilde{F}=\diag\{\tilde{c}_a\} \hs, \eeq
  with
  \beql{eta} \tilde{c}_a=+\sqrt{|c_a|^2+\eta} \hsc c_a\in\CC 
     \hsc a=1\ldots N:=\min(N_L,N_R) \eeq
  (for $N_L\neq N_R$ there is also a block of zeros with an appropriate 
  dimension).
\end{itemize}
The space of inequivalent $(F,\tilde{F})$ configurations is, therefore, $N$
(complex) dimensional. It can be parameterized by%
\footnote{More precisely, configurations $\{c_a\}$ related by a permutation
  are gauge equivalent, so the moduli space is obtained by dividing the 
  $\{c_a\}$ space by the permutation group.}
$\{c_a\}$ or, equivalently, by $\{b_a\}$, where
\beq b_a=c_a\tilde{c}_a \eeq
are the $N$ common eigenvalues of $\tilde{F}F$ and $F\tilde{F}$ 
(the extra eigenvalues, existing for $N_L\neq N_R$, all vanish).
As usual \cite{Moduli}, the moduli space can be parameterized also by 
gauge-invariant polynomials  
\[ T_l:=\tr(\tilde{F}F)^l \hsc l=1\ldots N \hs. \]
The gauge choice (\ref{fix}) also shows explicitly that the gauge symmetry 
unbroken by $F,\tilde{F}$ is generically $U(1)_D^N$ (diagonally embedded in 
$U(N_L)\otimes U(N_R)$). We adopt this gauge choice in the following classical
analysis.

We now turn to the F-term equations $dW=0$. We assume $\kp\neq0$ (for both 
left and right, as is the case for $N=2$ supersymmetry) and, to simplify the
analysis we ``absorb'' $\kp$ into a redefinition of $A$, $\xi$, $\mu$ and
$\lm$. Practically this means that we set $\kp=1$ in the superpotential%
\footnote{$\kp$ does not disappear! Rather it moves to the kinetic term of
  $A$.}
and, later, $\kp$ can be recovered by
\beql{kp} A\goto\kp A \hsc \xi\goto\xi/\kp \hsc \mu\goto\mu/\kp^2 
\hsc \lm\goto\lm/\kp \hs. \eeq
In the same way one can absorb $\lm$ in $\tilde{Q}$ (assuming it is
invertible), by setting $\lm=1$ (after absorbing $\kp$) and recover it
(before recovering $\kp$) by%
\footnote{$\kp$ and $\lm$ can also be recovered using symmetries. This is
explained in Appendix \ref{sym}.}
\beql{lm} Q\goto Q\lm \hsc m\goto \lm\inv m \hs. \eeq

The equations obtained are
\begin{eqnarray}
  \label{eom-F}   0 & = & m_FF+A_LF+FA_R \hs, \\ 
  \label{eom-TF}  0 & = & m_F\tilde{F}+\tilde{F}A_L+A_R\tilde{F} \hs, \\
  \label{eom-AL}  0 & = & \xi_LI_{N_L}+\mu_LA_L+F\tilde{F} \hs, \\ 
  \label{eom-AR}  0 & = & \xi_RI_{N_R}+\mu_RA_R+\tilde{F}F \hs. 
\end{eqnarray}
We now solve them for different choices of parameters.

\subsubsection{\boldmath $\mu_L\mu_R=0$}

Assume $\mu_R=0$. From eq. (\ref{eom-AL}) we obtain
\beq \tilde{F}F=-\xi_RI_{N_R} \hs, \eeq
which implies that all $c_a$'s are determined by $\xi_R$, they are equal and
if they do not vanish, then $N_L\ge N_R$. 

If $\mu_L$ also vanishes then a non-trivial value for $F$ (and, therefore,
also for $\xi_{L,R}$) is possible only for $N_L=N_R$ and then 
$\xi_L=\xi_R=:\xi$. Therefore, we have two possible situations:

\begin{description}
\item{\bf Type 1:} 
  $\mu_L=\mu_R=0$, $\eta\neq0$ and/or $\xi\neq0$ ($N_L=N_R=:N$)

  $\tilde{F}$ is a non-vanishing multiple of the identity, therefore, eq.
  (\ref{eom-TF}) leads to
  \beq m_F+A_L+A_R=0 \hs. \eeq
  The vacua can be parameterized by $A_L$, (which can be diagonalized by a
  $U(N)_D$ color rotation preserving $F$ and $\tilde{F}$). The gauge symmetry
  is broken, generically, to $U(1)_D^N$ (diagonally embedded in 
  $U(N)_L\otimes U(N)_R$).

\item{\bf Type 2:} $\mu_L=\mu_R=0$, $\eta=\xi=0$

  $F$ and $\tilde{F}$ vanish, so $A_L$ and $A_R$ are independent and can be
  diagonalized simultaneously by a $U(N_L)\otimes U(N_R)$ color rotation.
  The vacua are parameterized by $A_L$ and $A_R$ and the gauge group is 
  generically broken to $U(1)^{N_L}\otimes U(1)^{N_R}$.

\end{description}
If $\mu_L$ does not vanish%
\footnote{The case of $\mu_L=0\neq\mu_R$ is treated similarly, with identical
  results (exchanging $L \leftrightarrow R$ but {\em not} 
  $F\leftrightarrow\tilde{F}$).}
then $A_L$ is fixed by its equation of motion (\ref{eom-AL}): 
\beq A_L=-\frac{1}{\mu_L}(\xi_LI_{N_L}+F\tilde{F}) \eeq
and we have also two possible situations:

\begin{description}
\item{\bf Type 3:} $\mu_L\neq0=\mu_R$, 
  $\eta\neq0$ ($N_L=N_R$) and/or $\xi_R\neq0$ ($N_L\ge N_R$)

  $\tilde{F}$ does not vanish (and, moreover, has maximal rank), therefore eq.
  (\ref{eom-TF}) leads to
  \beq A_R=\left[\frac{1}{\mu_L}(\xi_L-\xi_R)-m_F\right]I_{N_R} \hs. \eeq
  Thus, in this case, there is a unique vacuum, in
  which the gauge symmetry is broken to $U(N_L-N_R)_L\otimes U(N_R)_D$ (the
  first factor is a subgroup of $U(N_L)$ and the second is diagonally 
  embedded in $U(N_L)\otimes U(N_R)$).

\item{\bf Type 4:} $\mu_L\neq0=\mu_R$, $\eta=\xi_R=0$

  $F$ and $\tilde{F}$ vanish, so $A_R$ is free and can be diagonalized by a
  $U(N_R)$ color rotation. It parameterizes the space of vacua and the gauge
  group is generically broken to $U(N_L)\otimes U(1)^{N_R}$.
\end{description}

\subsubsection{\boldmath $\mu_L\mu_R\neq0$}

If $\mu_L$ and $\mu_R$ do not vanish then both $A_L$ and $A_R$ are fixed by
their equations of motion
\beql{ALR}
   A_L=-\frac{1}{\mu_L}(\xi_LI_{N_L}+F\tilde{F}) \hsc
   A_R=-\frac{1}{\mu_L}(\xi_RI_{N_R}+\tilde{F}F) \hs.
\eeq
Observe that the above equations imply that the unbroken gauge symmetry is
the subgroup preserving $F$ and $\tilde{F}$ ($A_L$ and $A_R$ do not break
the gauge symmetry further). Eq. (\ref{eom-TF}) becomes
\beql{ctc} 
  \left(\hat{m}_F-\frac{1}{\mu}\tilde{c}_ac_a\right)\tilde{c}_a=0 \hs,
\eeq
where
\[ \hat{m}_F:=m_F-\left(\frac{\xi_L}{\mu_L}+\frac{\xi_R}{\mu_R}\right) \hsc
   \frac{1}{\mu}:=\frac{1}{\mu_L}+\frac{1}{\mu_R} \]
and this leads to two possible situations:

\begin{description}
\item{\bf Type 5:} $\hat{m}_F\neq0$ and/or $\frac{1}{\mu}\neq0$

  In this case there is at most one non-vanishing solution $c$ to eq.
  (\ref{ctc}), which means that $F\tilde{F}$ and $\tilde{F}F$ have (at most) 
  one non-vanishing eigenvalue $b$ (determined uniquely by the parameters),
  with some multiplicity $r$ ($0\le r\le N:=\min(N_L,N_R)$). Therefore, in
  this situation there is a finite number of discrete vacua, parameterized by
  $r$. The gauge group is broken to 
  $U(N_L-r)_L\otimes U(r)_D\otimes U(N_R-r)_R$. The range of values that $r$
  may assume depends on the parameters:

  \begin{description}
  \item{\boldmath $\eta\neq0$} ($N_L=N_R=N$)

    This leads to $\tilde{c}\neq0$, therefore, eq. (\ref{ctc}) implies that
    $r=N$ (a unique vacuum) and, moreover, if $\frac{1}{\mu}=0\neq\hat{m}_F$
    then there is no vacuum at all.

  \item{\boldmath $\frac{1}{\mu}=0\neq\hat{m}_F$} ($\eta=0$)

    This leads to $\tilde{c}=0$, which means $r=0$ (a unique vacuum).

  \item{\boldmath $\frac{1}{\mu}\neq0,\hs\eta=0$}

    In this case all values of $r$ are allowed and there are $N+1$ discrete
    vacua.
  \end{description}
        
\item{\bf Type 6:} $\hat{m}_F=\frac{1}{\mu}=0$

  In this case $F$ is unconstrained by eq. (\ref{ctc}) and it parameterizes the
  space of vacua. The gauge symmetry is broken, generically, to
  $U(N_L-r)_L\otimes U(1)_D^r\otimes U(N_R-r)_R$, where $r$ is the rank of
  $\tilde{F}$.
\end{description}

\subsection{The Classical \boldmath $SU(N_L)\otimes SU(N_R)$ Model}

We will also have to consider a variant of the above model, obtained by 
{\em freezing} the
$N=2$ vector multiplets corresponding to the two $U(1)$ factors
in the gauge group. By this we mean that we replace each such $N=2$ dynamical
superfield by a fixed constant (which, by supersymmetry, is also independent 
of the Fermionic coordinates and, therefore, contributes only to the scalar
component). The result is a model with an $SU(N_L)\otimes SU(N_R)$ gauge 
group, and almost the same matter content, the only difference being that
$\tr A$ is not a dynamical field, but rather a fixed parameter.
We will now list the differences between the models:
\begin{itemize}
\item The modulus $\tr\vev{A}$ becomes a parameter.

  One can absorb $\tr A$ in the mass parameters $m_L,m_R$ and $m_F$
  ($\xi$ is shifted too, but see the next paragraph), so we can assume 
   $\tr A=0$.

\item In the superpotential, $\xi$ multiplies a constant and, therefore, could
  be ignored. But we re-introduce these terms with a different interpretation:
  $\xi$ becomes a Lagrange multiplier, enforcing the constraint $\tr A=0$. 
  So we obtain the same equations (\ref{eom-F}-\ref{eom-AR}), the only
  difference being that we have also the above constraint, which is an
  additional equation, leading to the determination of $\xi$.

\item The FI terms disappear (since there are no $U(1)$ gauge fields). 
  Nevertheless, the D-term equations (\ref{D-eq}) remain the same, but with a
  different meaning: $\eta_L,\eta_R$ are no longer (FI) parameters but rather
  functions of the moduli -- reflecting the
  requirement that the other terms in the equation should be orthogonal to the 
  (traceless) generators of the gauge group and, therefore, must combine to a 
  multiple of the identity matrix.

  Observe that $\xi$ and $\eta$ undergo the same type of change
  (parameter $\goto$ modulus). This is obviously related to the fact that
  they are in the same $SU(2)_R$ multiplet.

\item For $Q=0$, $A$, $F$ and $\tilde{F}$ can still be diagonalized 
  simultaneously, but this time (compare with eq. (\ref{eta}))
  \beql{al} \tilde{c}_a=e^{i\al}\sqrt{|c_a|^2+\eta} \hs, \eeq
  where $\al,\eta\in\RR$. For $N_L=N_R=:N$, $\eta e^{i\al}$ is an additional 
  complex modulus, so the space of 
  inequivalent ($F,\tilde{F}$) configurations is $N+1$ (complex) dimensional.
  This change in dimension is reflected 
  by the existence of additional gauge invariant polynomials:
  \[ D:=\det F \hsc \tilde{D}:=\det\tilde{F} \]
  ($T_N\equiv\tr(\tilde{F}F)^N$ is a polynomial function of $D,\tilde{D}$ and
  $T_l$ with $l<N$, so there are indeed $N+1$ independent complex moduli).
\draftnote{For $N_L\neq N_R$, $\eta$ is forced to vanish, but $\al$ is still
    arbitrary. \newline
    HOW IS $\al$ REFLECTED IN THE INVARIANT POLYNOMIALS?}
\end{itemize} 
The above changes lead obviously to a different space of parameters and moduli.
To find it, we can use the same equations (\ref{eom-F}-\ref{eom-AR}),
with a modified meaning: tr$A$ vanishes identically and, therefore, the trace
of eqs. (\ref{eom-AL},\ref{eom-AR}) determines $\xi$:
\beq T_1\equiv\tr\tilde{F}F=-N_L\xi_L=-N_R\xi_R \hs. \eeq

\subsubsection{\boldmath $\mu_L=\mu_R=0$}
\secl{mu0}

Eqs. (\ref{eom-AL},\ref{eom-AR}) imply that both $\tilde{F}F$ and $F\tilde{F}$
are proportional to the identity. Therefore, for $N_L\neq N_R$,
$F$ and $\tilde{F}$ must vanish and we have the same moduli space as for the
previous model (type 2), parameterized by independent $A_L$ and $A_R$
(which, however, represent now less degrees of freedom, since they are
traceless). We will call this branch of vacua the Coulomb branch.
For $N_L=N_R$, $F$ and $\tilde{F}$ do not have to vanish and, for $m_F=0$
we obtain an 
additional branch, parameterized by $A_L=-A_R$, $D$ and $\tilde{D}$
(for $m_F\neq0$, $F$ and $\tilde{F}$ are forced to vanish by eqs.
(\ref{eom-F},\ref{eom-TF})).
This branch will be called the mixed branch.

\subsubsection{\boldmath $\mu_L\neq0=\mu_R$}
\secl{mu-yn}

Eq. (\ref{eom-AR}) implies that $\tilde{F}F$ is a multiple of the identity
(which means that it must vanish for $N_L<N_R$) and eq. (\ref{eom-AL}) leads to
\[ A_L=-\frac{1}{\mu_L}\left(F\tilde{F}-I_{N_L}\frac{1}{N_L}T_1\right) \hs. \]
Therefore, for $N_L<N_R$ we have a unique branch (again, as in the previous
model -- type 4),
parameterized by $A_R$ (with vanishing $A_L,F$ and $\tilde{F}$),
but for $N_L\ge N_R$ there may be another branch, with
non-vanishing $F$ and/or $\tilde{F}$ (rank $N_R$). In this case equations 
(\ref{eom-F},\ref{eom-TF}) imply
\[ \mu_Lm_F=\left(\frac{1}{N_R}-\frac{1}{N_L}\right)T_1 \hsc A_R=0 \hs, \]
so the situation is as follows: for $N_L=N_R$ there is an additional branch
if and only if $m_F=0$ and then it is parameterized by $D$ and $\tilde{D}$
($A_L$ vanishes). 
For $N_L>N_R$, $\tilde{F}F$ is fixed by the parameters, but the overall phase
of each of them ($\al$ in eq. (\ref{al})) remains free, so we have a compact
one ({\em real}) dimensional branch of vacua.
\draftnote{THIS IS STRANGE! WHAT IS THE GAUGE INVARIANT POLYNOMIAL
  PARAMETERIZING THIS BRANCH?}

\subsubsection{\boldmath $\mu_L\mu_R\neq0$}

When $\mu_L,\mu_R$ both do not vanish, $A_L$ and $A_R$ are determined by their
equations of motion (as in (\ref{ALR}))
\beql{SALR}
A_L=-\frac{1}{\mu_L}\left(F\tilde{F}-I_{N_L}\frac{1}{N_L}T_1\right) \hsc
A_R=-\frac{1}{\mu_R}\left(\tilde{F}F-I_{N_R}\frac{1}{N_R}T_1\right)
\eeq
and then eq. (\ref{eom-TF}) obtains the form
\beql{sc}
  0=\left(m_F+\frac{T_1}{\tilde{\mu}}-\frac{\tilde{c_a}c_a}{\mu}\right)
    \tilde{c_a} \hsc
\eeq
where
\[ \frac{1}{\mu}:=\frac{1}{\mu_L}+\frac{1}{\mu_R} \hsc
   \frac{1}{\tilde{\mu}}:=\frac{1}{N_L\mu_L}+\frac{1}{N_R\mu_R} \hs. \]
We now specialize to the case $N_L=N_R:=N$.
In this case, eq. (\ref{sc}) simplifies to
\beql{sce}   
0=\left[m_F+\frac{1}{N\mu}(T_1-N\tilde{c}_ac_a)\right]\tilde{c}_a \hs.
\eeq
For $\rec{\mu}\neq0$, there is at most one non-zero value $\tilde{c}c$ for
$\tilde{c_a}c_a$, so eq. (\ref{sce}) simplifies further to
\[ \mu m_F=\frac{N-r}{N}\tilde{c}c \hs, \]
where $r$ is the multiplicity of $\tilde{c}c$. This leads to the following
solutions:
\begin{description}
\item {\boldmath $\frac{1}{\mu}=0=m_F:$} $F$ and $\tilde{F}$ are free;
\item {\boldmath $\frac{1}{\mu}=0\neq m_F:$} there is a unique vacuum
  $F=\tilde{F}=0$;
\item {\boldmath $\frac{1}{\mu}\neq0=m_F:$} $F$ and $\tilde{F}$ are
  multiples of the identity, so there is a continuous moduli space, 
  parameterized by $D,\tilde{D}$;
\item {\boldmath $\frac{1}{\mu}\neq0\neq m_F:$} there are $N$ discrete vacua,
  parameterized by $r=0\ldots N-1$.
\end{description} 
This is quite similar to the moduli space of the previous model, the main
difference being the existence of a continuous moduli space of vacua also for 
$\frac{1}{\mu}\neq0=m_F$.

\subsection{The Quantum $SU(N_L)\otimes SU(N_R)$ Model}
\secl{FT-quant}

In this section we discuss modifications in the moduli space of the
$SU(N_L)\otimes SU(N_R)$ models caused by quantum corrections. As in the
classical analysis, we restrict our attention to vacua with $Q=\tilde{Q}=0$.

\subsubsection{\boldmath $\mu_L=\mu_R=0$}

When the model
has $N=2$ supersymmetry (\ie, $\mu=0$ and $\kp=\lm=\sqrt{2}$), the quantum
corrections are severely restricted: both branches described in subsection
\ref{mu0} survive, the ``mixed'' branch retains the structure of a direct
product and only the Coulomb parts are modified by quantum corrections%
\footnote{This follows from the arguments presented in \cite{APS9603}.}.
Even when the Yukawa couplings are modified, so that the $N=2$ supersymmetry
is explicitly broken, the Coulomb branch is not lifted.
To see this, one considers the $R$-symmetry with $R_Q=R_F=1$ and $R_A=0$.
It is anomaly-free (\ie, $R_{\Lm_L}=R_{\Lm_R}=0$) and is
respected by the classical superpotential, therefore, it must be respected
also by the dynamically generated superpotential. This means that the
full low energy effective superpotential is quadratic in $F$ and $Q$, and
the Coulomb branch (with $F=Q=0$; parameterized by $A$) is not lifted by
quantum corrections.

\subsubsection{\boldmath $N_L=N_R:=N$, $\mu_L=-\mu_R:=-\mu\neq0$}

In this case, at scales below $\mu$, $A_L$ and $A_R$ are expected to 
decouple. When $\mu\gg\Lm_L,\Lm_R$, the decoupling occurs at weak gauge 
coupling and can be analyzed semi-classically. Integrating out $A_L,A_R$ by
their equations of motion (see eqs. (\ref{SALR}) and recall that we use the 
gauge choice (\ref{fix}), in which $F$ and $\tilde{F}$ commute)
\beq A_L=-A_R=B/\mu \hsc B:=F\tilde{F}-\frac{1}{N}I_N\tr(F\tilde{F}) \eeq
leads to the effective superpotential%
\footnote{Recall that we chose $\lm=\kp=1$. Other values for them
  can be incorporated
  by the transformations (\ref{kp},\ref{lm}). In particular, the condition
  for the existence of this branch is $\mu_L/\kp_L^2=-\mu_R/\kp_R^2$.}
\beql{Weff}
  W = \tr(m_L\tilde{Q}_LQ_L)+\tr(m_R\tilde{Q}_RQ_R)
    +\frac{1}{\mu}[\tr(\tilde{Q}_LBQ_L)-\tr(\tilde{Q}_RBQ_R)] \hs,
\eeq
so the resulting effective model, without adjoint fields, will have
the tree-level superpotential (\ref{Weff}). This branch is also not lifted by
quantum corrections, since it is protected by the non-anomalous $R$ symmetry
with $R(Q)=1$ and $R(F)=0$.

\subsubsection{\boldmath $\mu_L\neq0=\mu_R$}

Finally, we discuss the situation, when only one of the adjoint fields --
$A_L$ -- is massive%
\footnote{We are grateful to
David Kutasov for a discussion on this subsection.}.
Consider, for simplicity, a model without fundamentals ($n_L=n_R=0$).
Integrating out $A_L$ (assuming, as before, $\mu_L\gg\Lm_L,\Lm_R$)
\[ A_L=-\frac{1}{\mu_L}(F\tilde{F}-\frac{1}{N_L}\tr F\tilde{F}) \]
leads, at scales below $\mu_L$, to a model with a tree level superpotential
\beq \Wtree=-\frac{1}{2\mu_L}[\tr(M^2)-(\tr M)^2]+\tr[(m_F+A_R)M] \hsc
     M:=\tilde{F}F \hs,
\eeq
so the classical model in this case
has a branch of vacua with $M=0$, parameterized by $A_R$.
Unlike the previous cases, here symmetry does not protect this branch from
the dynamical generation of an $A_R$-dependent superpotential that would lift
it. In fact we know that such a superpotential is generated in a closely
related model. Taking $\Lm_R=0$ and $\frac{1}{\mu_L}=0$, one obtains a model
with gauge group $SU(N_L)$, under which $F,\tilde{F}$ transform as $N_R$
fundamentals and antifundamentals
and $A_R$ transforms as $N_R^2-1$ singlets. For $N_L<N_R$ this is {\em almost}
``magnetic SQCD'' of \cite{Seiberg9411}. In the latter model, there are
$N_R^2$ singlets $\hat{A}_R$, coupled to the meson $M$ by a superpotential
tr$\hat{A}_RM$, so the only difference between the models is that
one of the singlets of magnetic QCD (tr$\hat{A}_R$)
is ``frozen'' here, becoming a
parameter $m_F$. As is known, quantum effects indeed generate an
$\hat{A}_R$-dependent superpotential in this model. 
In section \ref{M-th}, we will show, using M theory, that
this also happens for all the present models. We will return to the field
theoretical consequences in section \ref{disc}.
The explicit determination of the generated superpotential is left for
future work.

\internote{QUANTUM LIFTING OF BRANCHES OF VACUA:
 \nl For $\mu\neq0$ a dynamically generated superpotential is expected at
 scales below $\mu$ (generalization of \cite{IS9509}; see \cite{HOO9706}).
 \nl Expectations (from string considerations) for surviving vacua:
 \nl $N_L=N_R$:
 \nlb $\mu_L\neq\mu_R$, $m_F=0$ ($D,\tilde{D}$): all;
 \nl $N_L\neq N_R$:
 \nlb $\mu_L\mu_R\neq0$, $\mu_L\neq\mu_R$ (generically $r,\al$):
 generically none;
 \nlb $\mu_L=\mu_R\neq0$ ($F$ mod $T_1$): generically none.}

\newsection{The Type IIA Brane Configuration}
\secl{IIA}

Our next step is to construct a configuration of branes in Type IIA string
theory that will reproduce the results obtained in the previous section.

\subsection{The Classical Description}

We will consider (classically) flat branes in flat 10 dimensional spacetime:
solitonic fivebranes
(Denoted $\NS$) and Dirichlet 4-branes and 6-branes (denoted $D_4$ and $D_6$
respectively). The $D_4$ branes will have boundaries on $\NS$ branes.
The orientation of branes is described in the following table
\beq\begin{array}{|c||c|c|c|c|c|}
  \hline \rule[-1.3ex]{0em}{4ex}
  {\rm Type} & x^\mu,\;\mu=0,1,2,3 & v:=x^4+ix^5 & s=(x^6+ix^{10})/R_{10} &
         x^7 & w:=x^8+ix^9 \\
  \hline
  \NS & \mbox{---} & \mbox{---} &        \bullet & \bullet & \bullet \\    
  D_4  & \mbox{---} & \bullet    &  [\mbox{---}]  & \bullet & \bullet \\
  D_6  & \mbox{---} & \bullet    &    \bullet & \mbox{---} & \mbox{---}  \\
  \hline\end{array}
\eeq
(we will also have $\NS$ and $D_6$ branes rotated in the
  $(v,w)$ space; they will be described shortly).
In the table, a dash `---' represents a direction along which the brane is 
extended and a bullet `$\bullet$' represents a direction transverse to the 
brane. For the $D_4$ brane, `[---]' means that the brane does not extend 
along the full $x^6$ axis, since at least one of its sides ends on a 
$\NS$ brane. We will have both ``finite'' and ``semi-infinite'' $D_4$ branes
in the $x^6$ direction. 

In the next section we will reinterpret the type IIA string theory discussed
here as M theory on $\RR^{10}\times S^1$. We, therefore, incorporated in the 
table also the extension of the branes in the 11th (compact) dimension.
$x^{10}$ is the corresponding coordinate and $2\pi R_{10}$ is its periodicity.

The ``finite'' $D_4$ branes (\ie, those that are restricted from both sides
in the $x^6$ direction by a $\NS$ brane) are the only branes in the 
configuration that have a finite extent in the ``internal space''
($x^i$, $i=4\ldots9$), therefore, at a low enough energy, the dynamics of the 
configuration is approximately the dynamics of the finite $D_4$ branes in the
background of a fixed configuration of all other branes -- an effective field
theory on the world volume of the finite $D_4$ branes. Because of the finite
extent in the $x^6$ direction, this dynamics will be effectively 4 
dimensional. 

A collection of branes of the above types preserves $\rec{4}$ of the original
10 dimensional $N=2$ supersymmetry \cite{HW9611}, therefore, the corresponding
effective 4 dimensional field theory will have $N=2$ supersymmetry. 
A $\NS$ brane with
a different orientation then that described above will break generically all
the remaining supersymmetry. However there is a way to rotate a $\NS$ brane
in such a way that half of the remaining supersymmetry is preserved
and the resulting 4 dimensional field theory has
$N=1$ supersymmetry. In particular, one can perform the following rotation:
\beql{rot}
  \left(\begin{array}{c}v\\w\end{array}\right)\goto
  \left(\begin{array}{rr}\cos\th&-\sin\th\\ \sin\th&\cos\th\end{array}\right)
  \left(\begin{array}{c}v\\w\end{array}\right) \hs.
\eeq
Indeed, this rotation can be described as an $SU(2)$ transformation:
\[ \left(\begin{array}{c}x^4+ix^8\\x^5-ix^9\end{array}\right)\goto
   \left(\begin{array}{cc}e^{i\th}&\\&e^{-i\th}\end{array}\right)
   \left(\begin{array}{c}x^4+ix^8\\x^5-ix^9\end{array}\right) \]
and this guarantees \cite{BDL9606} the preservation of $N=1$ supersymmetry.
Obviously, $w$ in (\ref{rot}) can be replaced by $e^{i\vph}w$, $\vph\in\RR$
(or equivalently $v\goto e^{-i\vph}v$), leading to a different rotation, so
a rotation of this kind is parameterized by two real angles. 
Identical considerations lead to the possibility to rotate a $D_6$ brane,
using the transformation (\ref{rot}). In fact, one can use the transformation
(\ref{rot}) to rotate an arbitrary number of $\NS$ branes and $D_6$ branes
(each with different rotation parameters) without spoiling $N=1$ 
supersymmetry.

At this stage we can specify the precise configuration of branes that will be
considered:
\begin{itemize}
\item Three $\NS$ branes, denoted by $S_L,S_M,S_R$, according to their order
  along the $x^6$ axis. We choose $S_M$ to be parallel to the $v$ plane, while
  the other two are rotated with rotation parameters $(\th_L,\vph_L)$ and
  $(-\th_R,\vph_R$) (note the sign difference).
\item $(n_L,N_L,N_R,n_R)$ $D_4$ branes extended in the $x^6$ direction in the
  intervals 
  \[-\infty\leftrightarrow S_L\leftrightarrow S_M\leftrightarrow S_R
    \leftrightarrow+\infty \]
  respectively.
\end{itemize}

\subsection{Identification of the Effective Field Theory}
\secl{Identify}

\begin{figure}
\pct{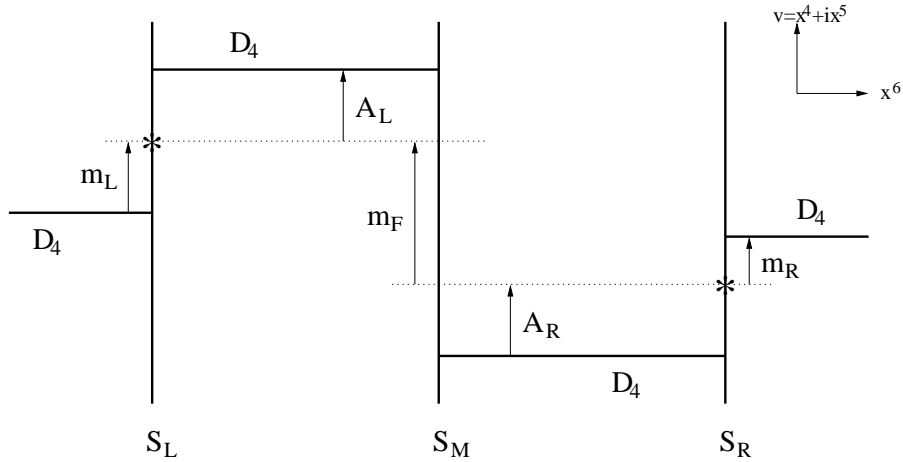}
\caption{Projection on the $(v,s)$ subspace. For brevity, only one $D_4$ brane
is displayed in each region; the displacements are the eigenvalues of the
corresponding matrices ($m_L,A_L,A_R,m_R$).}
\label{vs-diag}
\end{figure}

\begin{figure}
\pct{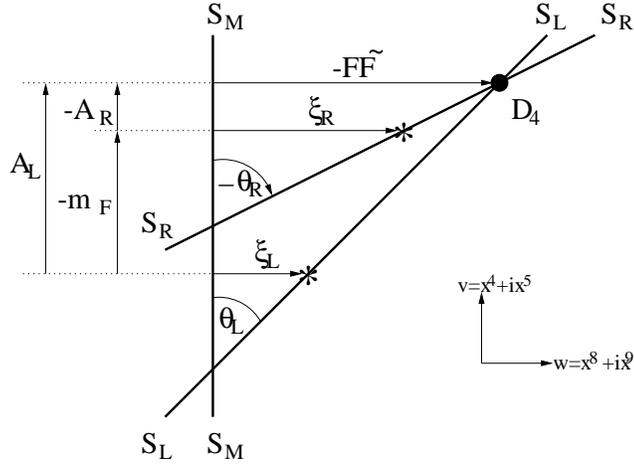}
\caption{Projection on the $(v,w)$ subspace. The displacements described are
for an $S_L\leftrightarrow S_R$ $D_4$ brane (denoted by a bullet $\bullet$).}
\label{vw-diag}
\end{figure}

We argue that the effective field theory on the $D_4$ branes is indeed the one
presented in section \ref{FT}. Moreover, we suggest
a quantitative identification between most of the parameters and
moduli of the field theory and the parameters determining the brane
configuration (see figures \ref{vs-diag},\ref{vw-diag})%
\footnote{Parts of this identification already appeared in the literature. We
  combine them with some new elements and obtain an almost complete picture.}
\footnote{In the following identification we assume unit string tension
$T_\st$. $T_\st$ can be recovered by dimensional analysis. For example,
$\Dl v$ represents $T_\st\Dl v$, which is the mass of a string of length
$\Dl v$. $w$ represents $T_\st^{3/2}w$ and, therefore, tan$\th$ represents
$T_\st^{1/2}\tan\th$.}.
To specify the location of $S_L$ and $S_R$ in the $v$ and $w$ planes, we 
{\em choose} a reference point on each brane (it is denoted by star
``*'' in the figures) and refer to its coordinates.
The arbitrariness of this choice will be discussed below. We now list the
identification. The parameters are
\begin{itemize} 
\item $v$-position: (here the $D_4$ branes are the semi-infinite ones)  
   \[ m_F = (S_L\mbox{ brane})-(S_R\mbox{ brane}) \hs, \]
   \[ -m_L = (D_{4L}\mbox{ brane})-(S_L\mbox{ brane}) \hsc 
       m_R = (D_{4R}\mbox{ brane})-(S_R\mbox{ brane}) \hs; \] 
\item $w$-position: 
   \[ \xi_L = (S_L\mbox{ brane})-(S_M\mbox{ brane}) \hsc
      \xi_R = (S_R\mbox{ brane})-(S_M\mbox{ brane}) \hs; \]
\item $s$-position: ($\tau=\frac{\th}{2\pi}+i\frac{4\pi}{g^2}$)
   \[ i\tau_L = (S_L\mbox{ brane})-(S_M\mbox{ brane}) \hsc
      -i\tau_R = (S_R\mbox{ brane})-(S_M\mbox{ brane}) \hs; \]
\item orientation of the $\NS$ branes:
   \[ \mu_L=e^{i\vph_L}\tan\th_L \hsc \mu_R=e^{i\vph_R}\tan\th_R \]
\end{itemize}
and the moduli are (here the $D_4$ branes are the finite ones):
\begin{itemize}
\item $v$-position:  
  \[ A_L=(D_{4L}\mbox{ brane})-(S_L\mbox{ brane}) \hsc                
    -A_R=(D_{4R}\mbox{ brane})-(S_R\mbox{ brane}) \hs; \]               
\item $w$-position:
  \[ -F\tilde{F}=(D_{4L}\mbox{ brane})-(S_M\mbox{ brane}) \hsc                
     -\tilde{F}F=(D_{4R}\mbox{ brane})-(S_M\mbox{ brane}) \hs. \]
\end{itemize}
Also, the FI parameters $\eta_L$ and $\eta_R$ are related to the 
$x^7$-positions of $S_L$ and $S_R$, respectively, relative to $S_M$.
When the fundamentals are realized by $D_6$ branes (see below),
$\tilde{Q}Q$ is related to the $w$ positions of $D_4$ branes connecting
$D_6$ branes and 
the Yukawa couplings $\lm$ are related to the rotation of the $D_6$ branes.
The only other parameters that do not appear in the above list are the Yukawa
couplings $\kp$ of the bi-fundamentals.
Actually, in deriving the above relations, we used
the results of the previous section, where we set $\kp=\lm=1$. As explained
there, $\lm$ and $\kp$ can be recovered by using eqs. (\ref{lm}) and 
(\ref{kp}), so in the above list each field-theoretical quantity actually
represents the modified expression obtained after recovering $\lm$ and $\kp$.
For example, the orientation of the $\NS$ branes is related to 
$e^{i\vph}\tan\th=\mu/\kp^2$.
As long as we consider configurations that preserve $N=2$ supersymmetry, $\kp$
and $\lm$ are fixed, so nothing essential is missing. But when some
of the branes are rotated, $N=2$ supersymmetry is broken and the 
corresponding behavior of the Yukawa coupling may be important. In particular,
qualitative changes are expected when they vanish.
\internote{This may be discussed in section \ref{disc}.}

In the following we list the evidence for the above identification.

\vspace{5mm}\noindent
{\bf Field Content}

The field content was identified already in previous works 
\cite{HW9611,BH9704}. $N=2$ vector multiplets originate from
strings connecting two $D_4$ branes which extend between {\em the same} $\NS$
branes, while $N=2$ hypermultiplets originate from the other 4-4 strings.
The identification of the vector field content leads also to the 
identification of the gauge symmetry $U(N_L)\otimes U(N_R)$.

Note that we could replace the semi-infinite $D_4$ branes by $D_6$ branes,
obtaining the same matter content. For 
example, the $D_4$ branes extended to $+\infty$ can be replaced by $D_6$
branes located between $S_M$ and $S_R$ (along the $x^6$ axes). Indeed, when
such a $D_6$ brane is moved to the right and crosses $S_R$, a $D_4$ brane
is generated that extends between the $D_6$ and $S_R$ branes, so a
semi-infinite $D_4$ brane can be viewed as connected to a $D_6$ brane at 
infinity. When the crossing $\NS$ and $D_6$ branes are orthogonal to each 
other, this motion seems to be irrelevant in the effective field theory on the
$D_4$ branes%
\footnote{The relevance of this motion in other situations was discussed
in \cite{BH9704} and \cite{AH9704}.},
\internote{This may be discussed in section \ref{disc}.}
allowing to switch 
between these two representations. In most of the discussion it will not be
necessary to introduce $D_6$ branes%
\footnote{The $D_6$ branes are essential for the description of the Higgs
  branch of the field theory on the $D_4$ branes, but we do not consider this
  branch of vacua in the present work.}.

\vspace{5mm}\noindent
{\bf Parameters and Moduli}

There is a difference in the expected geometric realization of
field-theoretical parameters and moduli:
given the fields, their dynamics is dictated by the action, 
characterized by its parameters.
In the brane realization of the field theory, the dynamics is that of the
finite $D_4$ brane and it is dictated by the geometric configuration of the
infinite -- non-dynamical -- branes. Therefore, this is what the parameters of
the field theory should describe geometrically.
The action determines, in particular, the moduli space of vacua, and this 
corresponds to the possible geometric locations of the finite -- dynamical --
$D_4$ branes. Therefore the moduli of the field theory should determine
aspects of the geometric configuration that involve only the finite $D_4$
branes. 

This distinction between parameters and moduli is obeyed in the
identification described above.

\vspace{5mm}\noindent
{\bf Dimensionality and Redundancy}

One can verify that the dimensionality of the parameters and moduli indeed 
agrees with the one of the geometric quantities with which they are 
identified%
\footnote{Of course, only gauge invariant quantities can have a geometric
  meaning, so in $A_L,A_R,\tilde{F}F,F\tilde{F}$, only the eigenvalues should
  be considered. Moreover, in the identification of $m_L,m_R$, we assume that
  they are diagonal.}.
In this context one should observe that there is a redundancy in the
description
of both the field theory and the brane configuration. In the brane 
configuration this is the choice of a reference point on $S_L$ and $S_R$.
One can verify that a shift of this reference point
has exactly the same effect as a (scalar) shift of $A_L$ and $A_R$, 
respectively, accompanied by a change of parameters that leaves the action
invariant, up to an additive constant. A shift in $A_L$, for example, should
be accompanied by a shift of $m_F$ and $m_L$ and, if $\mu_L\neq0$, also of 
$\xi_L$. Such a shift can be used to set to zero some redundant parameters
(see footnote \ref{Redund}). All this is reflected exactly in the geometric
description.

\vspace{5mm}\noindent
{\bf The Equations of Motion and the Moduli Space of Vacua} 

\begin{figure}
\vspace{-5mm}
\pct{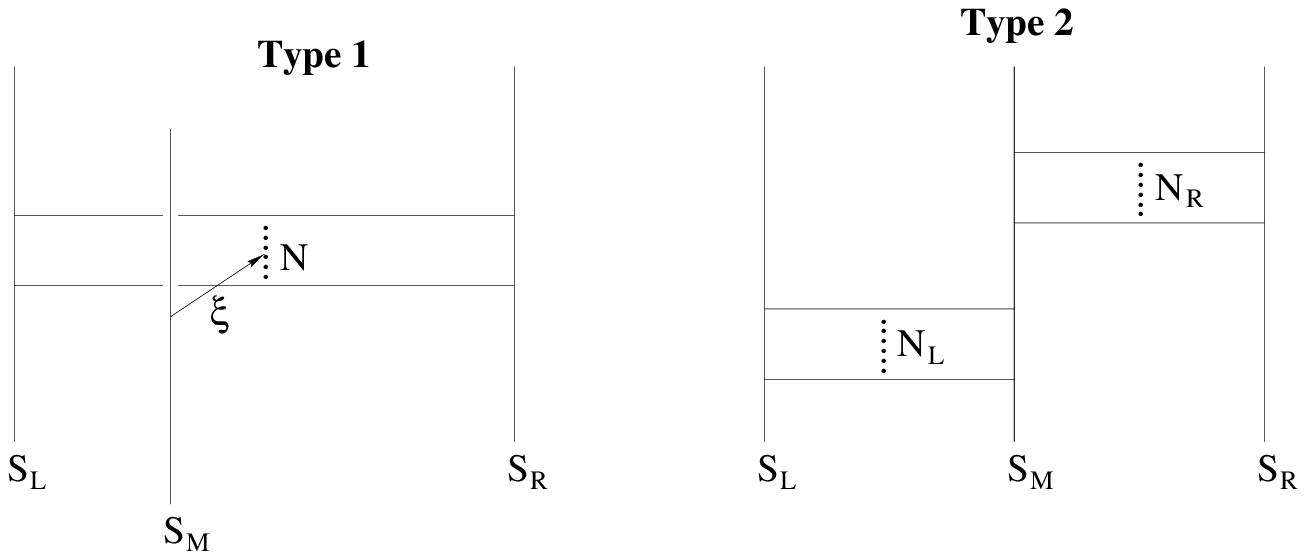}
\vspace{-1mm}
\pct{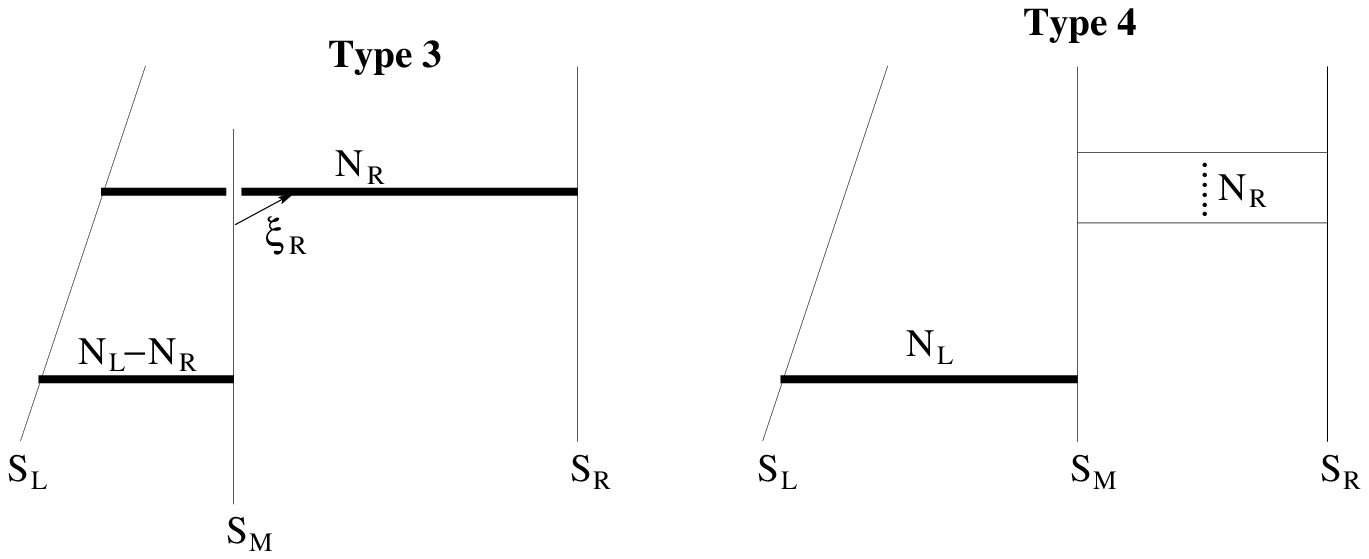}
\vspace{-1mm}
\pct{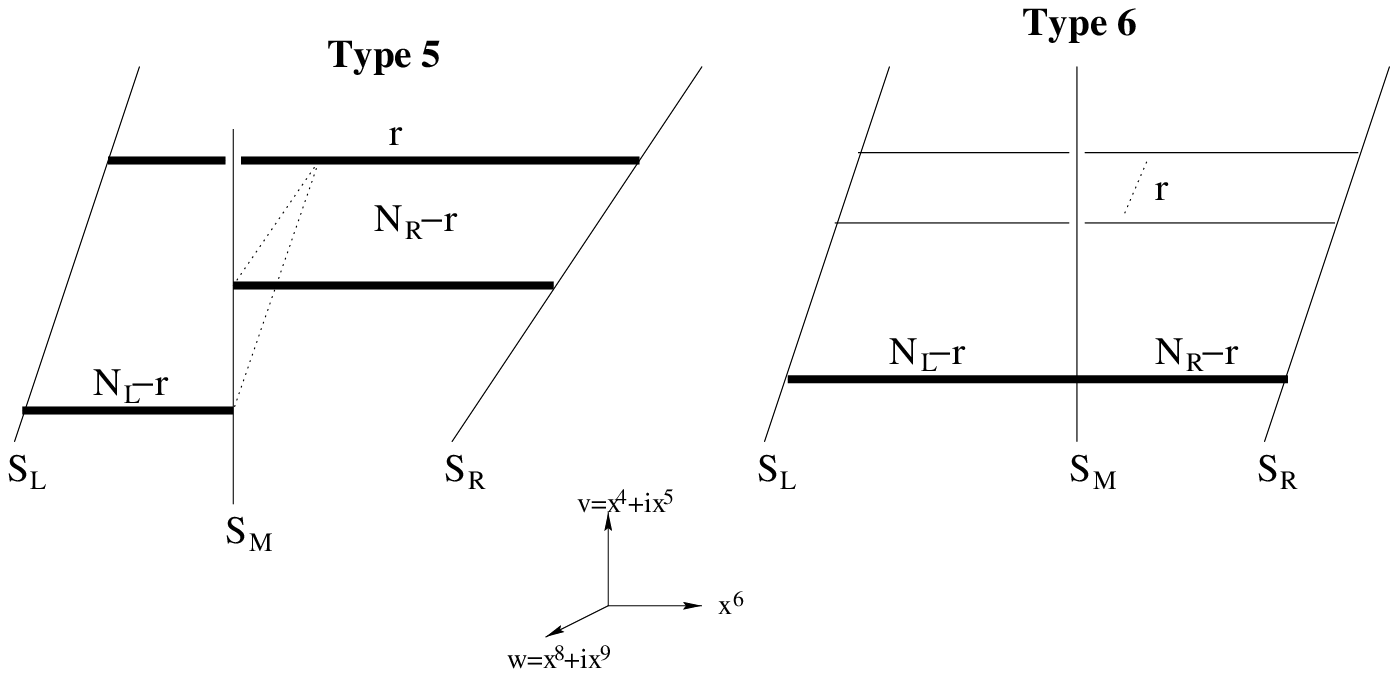}
\caption{The possible brane configurations. The types of configurations
correspond to the types of vacua found in section \ref{FT}.}
\label{mod-fig}
\end{figure}

The geometric identification implies naturally qualitative and quantitative
relations between the various parameters and moduli (these were already
used in the discussion of redundancy above). In the field theory, these
relations are the equations of motion derived from the action, therefore, the
identification of the relations from the geometry leads,
at least to a large extent, to the
identification of the action and, in particular, the superpotential.
With this in mind, it is worthwhile to identify the geometric origin of the
equations of motion:
\begin{itemize}
\item The restrictions on $\eta$, obtained from the D-term equations, follow
  directly from the requirement that the $D_4$ branes should have 
  $x^7=$const..
\item A non-zero eigenvalue for $\tilde{F}$ corresponds to an
  $S_L\leftrightarrow S_R$ $D_4$ brane, therefore, the left and right $D_4$
  branes from which it is formed should have the same $v$ and $w$
  (see fig. \ref{vw-diag}).
  The $w$ position implies that $\tilde{F}F$ and $F\tilde{F}$ must have
  common non-zero eigenvalues (which follows from the D-term equations) and 
  the $v$ position implies that for such an eigenvalue, $m_F+A_L+A_R=0$ 
  (which follows from eqs. (\ref{eom-F},\ref{eom-TF})).
\item From a diagram in the $(v,w)$ space (fig \ref{vw-diag})
  one obtains, for an $S_L\leftrightarrow S_R$ $D_4$ brane,
  $\xi+\mu A+\tilde{c}c=0$, which is precisely the content of eqs. 
  (\ref{eom-AL},\ref{eom-AR}). In particular, $A$ vanishes for $\mu=\infty$
  and is free for $\mu=0$.
\end{itemize}
Finally, one can classify all the possible brane configurations. The result
agrees completely with the field-theoretical analysis and one observes the 
six types of situations described in section \ref{FT} (see figure
\ref{mod-fig}).
Actually, once the equations of motion are identified geometrically, the
moduli spaces should obviously agree too.

\vspace{5mm}\noindent
{\bf\boldmath $N=2$ Supersymmetry}

In the brane configuration described above,
4 dimensional $N=2$ supersymmetry is broken iff the
$\NS$ branes are not all parallel. This is identified with non-vanishing masses
for the adjoints, which indeed break the supersymmetry to $N=1$. Changes in
the Yukawa couplings $\kp,\lm$ also break $N=2$ supersymmetry,
but these parameters are not represented geometrically in the above
configuration%
\footnote{As remarked above, when the fundamentals are realized by $D_6$ 
branes (and not by semi-infinite $D_4$ branes),
$\lm$ corresponds to the rotation of the $D_6$ branes, which indeed breaks the
$N=2$ supersymmetry.}.

\vspace{5mm}\noindent
{\bf R-Symmetry}

A rotation in the ``internal space'' (\ie, not acting on $x^\mu$,
$\mu=0,1,2,3$) rotates also $\th_\al$ -- the Fermionic coordinate of
the four dimensional superspace and, therefore, corresponds to an 
$R$-symmetry of the field-theoretical model.
The relations suggested above are compatible with the following identification:
$SU(2)_{789}$ (rotations in the (789) space) is the $SU(2)_R$ symmetry
and $U(1)_{45}$ and $U(1)_{89}$ (rotations in the $v$ plane and $w$ plane 
respectively) are the $U(1)_R$ symmetries characterized by the following
charges:
\beql{U1RT}\begin{array}{|c||c|c||c|c|c|c||c|c|c|c||c|c||c|}
  \hline \rule[-1.3ex]{0em}{4ex}
    &v &w &\th_\al& A&F,\tilde{F}&Q,\tilde{Q}&\xi&\mu&m_F&m&\kp&\lm&\Lm^b\\
  \hline
  R_{45}& 2&  & 1 & 2&           &           &   & -2&  2&2&   &   & 2b  \\
  R_{89}&  & 2& 1 &  &         1 &         1 & 2 &  2&   & &   &   &     \\
  \hline\end{array}\eeq
(the charges are left$\leftrightarrow$right symmetric, so we dropped the
$L,R$ subscripts). Observe that $\kp$ and $\lm$ are neutral, so the 
transformations (\ref{kp},\ref{lm}) do not affect the $R$ charges above.
For later convenience, we include in the table the charge of the
  instanton factor $\Lm^b$, representing the chiral anomaly. The one-loop
  beta function coefficients are $b_{L,R}=2N_{L,R}-(N_{R,L}+n_{L,R})$.

\supnote{More information about this subject is given in section \ref{U1R}.}

\subsection{Quantum effects}
\secl{IIA-quant}

So far, we considered the classical description of branes. There are, however,
quantum effects that change the situation described above considerably.

\subsubsection{The Bending of \boldmath $\NS$ Branes}
\secl{bend}

Classically, a $\NS$ brane is flat and the branes discussed in this work have
constant $s$. However, as observed in \cite{Witten9703}, when a $D_4$ brane
ends on a $\NS$ brane, the $\NS$ brane bends. For example, when a single $D_4$
brane at $v=0$ ends on the left side of a $\NS$ brane extending classically in
the $v$ direction, the $s$ coordinate of the $\NS$ brane behaves
asymptotically as $s=\log v$. One implication of this effect is that the
$U(1)_{45}$ symmetry of the classical $\NS$ brane is broken quantum
mechanically and this fits well with the $U(1)_R$ anomaly known in field 
theory \cite{EGKRS9704}. 
\supnote{One can formulate a general geometric formula for the anomalies ---
  see section \ref{Anomaly}.}
Another consequence of this bending, that is more important to us, is the 
freezing of the $N=2$ vector multiplets corresponding to the $U(1)$ gauge
factors. To understand this, one should observe that, since the $\NS$ branes
are bent, with the ``center of bending'' determined by the location of the
$D_4$ branes, a change in this location will be accompanied by a change in the
location of the $\NS$ brane, including the asymptotic parts, and it can be
shown \cite{Witten9703} that such a change will cost an infinite amount of
(kinetic) energy. Therefore, the modulus corresponding to the average location
of the $D_4$ branes is frozen and, by $N=2$ supersymmetry, this happens to the
whole $N=2$ vector multiplet to which it belongs. This argument was given for
an $N=2$ supersymmetric configuration, however, since this effect is seen to
be related to the local interaction between a $\NS$ brane and the $D_4$ branes 
ending on it, it is not expected to depend on an existence of other, remote,
branes that may break the $N=2$ supersymmetry.
This freezing means that, while classically we obtained an effective 
$U(N_L)\otimes U(N_R)$ field theory, the incorporation of the quantum effects
will lead to a realization of a
quantized $SU(N_L)\otimes SU(N_R)$ field theory.

\internote{We assume that there is no change in the superpotential.
  In particular, for $N_L=N_R=2$, terms proportional to $D$ and $\tilde{D}$
  could, in principle, appear in a renormalizable superpotential, but we
  assume they do not.}

\subsubsection{Inter-brane Forces} 
\secl{force}

Another quantum effect in brane dynamics was discovered in \cite{EGKRS9704}. 
Investigation of the moduli space of $N=1$ SQCD led to the observation that
$D_4$ branes between two {\em non-parallel} $\NS$ branes interact with each
other and with some other $D_4$ branes connected to the same $\NS$ branes.
Consider, for example, a $D_4$ brane between two $\NS$ branes $S_L,S_R$ which
are rotated with respect to each other by $90^\circ$.
If there is another $D_4$
brane between $S_L$ and a $D_6$ brane {\em parallel} to $S_L$, then the two
$D_4$ branes will be either attracted to each other (when they end on opposite
sides of $S_L$),
or repelled from each other (when they end on the same side).
This interaction has infinite range; it is a Coulomb-like interaction in the
co-dimension of the intersection of the $D_4$ brane with the $\NS$ brane.
Other $D_6$ branes can ``screen'' these forces in some situations, as described
in \cite{EGKRS9704}.
The effect of this interaction is to lift moduli spaces of 
vacua, when they exist, since vacua correspond to configurations in which all
branes are in equilibrium.
We will discuss brane forces in the present models in section \ref{disc}.
Before that, we will gain some relevant information from an analysis using
M-theory.

\newsection{Analysis in M Theory}
\secl{M-th}

In this section, we follow the approach initiated in \cite{Witten9703},
interpret the type IIA brane configuration described in the previous section
as a fivebrane in M theory and exploit the resulting simplifications to
obtain information about the corresponding field theory and also about
brane dynamics.

We start by describing how the description given in this section is related
to that of the previous section%
\footnote{We use here arguments given at \cite{Witten9703,Witten9706}.}.
There are three independent length scales relevant to this discussion: the
string 
length scale $l_\st$ (inversely related to the string tension), the length
$l_6$ of the (finite) $D_4$ branes and the radius $R_{10}$ of the 11th
dimension. In terms of these, the string coupling constant $g_\st$, the 
(inverse) gauge coupling constant Im$\tau$ and the 11 dimensional plank length
$l_{\rm pl}$ are given by
\beq g_\st=\frac{R_{10}}{l_\st} \hsc \im\tau=\frac{l_6}{R_{10}} \hsc
     l_{\rm pl}^3=l_\st^2R_{10} \hs. \eeq
The realization of the field theory by a type IIA brane configuration, as
described in the previous section, is performed in the limit of small $g_\st$,
(\ie\ $R_{10}\ll l_\st$), where string perturbation theory is reliable. One
also considers length scales above $l_\st$ (so that all the
massive modes of the string can be ignored) and above $l_6$
(so that the effective field theory is 4 dimensional). 
For a sufficiently small string coupling,
the gauge coupling is small at the string scale and
significant quantum effects in the gauge theory appear only at very low
energies, where complications of string theory, including gravitation,
are expected to be negligable. This leads to the conclusion that the dynamics
is reliably described by a gauge field theory or, in other words, that this
brane configuration indeed realizes a gauge field theory and, therefore, can
be used to investigate its properties.

In the present section, we consider a different limit:
$l_{\rm pl}\ll R_{10},l_6$.
In this range of parameters $g_\st$ is large and string perturbation theory is
not reliable. However, one can use instead the dual -- M theory --
description of the model.
If all the characteristic scales are large with respect to $l_{\rm pl}$
(as $R_{10}$ and $l_6$ are),
the low energy (long wavelength) approximation to $M$ theory is
expected to be sufficient.

In the passage between the two limits one keeps $\tau$ fixed,
so classically we keep considering the
same effective field theory. There is by now much evidence that also in the
quantum theory many aspects do not change (even when there is only $N=1$
supersymmetry \cite{HOO9706,Witten9706,BIKSY9706,HSZ9707,NOS9707%
,SS9708,CS9708,BO9708,AOT9708}).
In particular, this method is
expected to reproduce correctly the aspects that we will consider: the vacuum
structure and the low energy gauge coupling in a Coulomb phase.

\subsection{The Corresponding M Theory Configuration}

Both $\NS$ branes and $D_4$ branes of type IIA string theory, correspond to the
same object in M theory -- a fivebrane, therefore, the brane configuration
described in section \ref{IIA} corresponds, in M theory, to a single fivebrane,
with a non-trivial topology. The worldvolume of this fivebrane is of the form 
$\RR^{3+1}\times\Sg$, where $\RR^{3+1}$ is the 4 dimensional spacetime,
parametrized by $x^\mu$, $\mu=0,1,2,3$ and $\Sg$ is a two dimensional surface
in the internal space $\Sc$, parametrized by $v,w$ and $s$ 
(each connected component of the curve has fixed $x^7$, so this 
coordinate will not be important in the following). 
Moreover, to obtain (at least) $N=1$ supersymmetry in the 4 dimensional 
effective field
theory, $\Sg$ is required \cite{BBS9507,BBMOOY9608} to be a complex Riemann
surface in the
complex structure induced by the coordinetas $v,w,s$ of $\Sc$. This also 
implies that $\Sg$ is smooth and generically it will have no singularities,
therefore, for $l_{\rm pl}$ small enough, the low energy (long wavelength)
approximation of M theory is justified.

As explained in \cite{Witten9703} (extending results from \cite{Verlinde}),
the resulting low energy effective 4 dimensional field theory contains $g$
abelian gauge fields, where $g$ is the genus of $\Sg$, and $\Sg$ is the 
Seiberg-Witten curve for this theory, \ie, it determines the low energy
effective gauge coupling. The $N=2$ supersymmetric
brane configuration of section
\ref{IIA} leads to a fivebrane worldvolume with genus
$N_L+N_R-2$, supporting the claim that this is the low energy limit of an
$SU(N_L)\otimes SU(N_R)$ (and not of $U(N_L)\otimes U(N_R)$) gauge theory.

\subsection{The Curve With \boldmath $N=2$ Supersymmetry}

Ref. \cite{Witten9703} considers an $N=2$ supersymmetric version of the
model discussed here (corresponding to $\mu=0$ and $\kp=\lm=\sqrt{2}$ in the
present notation). The dependence of the corresponding curve on the 
coordinates $(v,t)$ was determined (by requiring an appropriate asymptotic
behavior) to be  
\beql{N=2-gen}
  C_LQ_L(v)t^3-P_L(v)t^2+P_R(v)t-C_RQ_R(v)=0 \hsc t=t_0e^{-s} \hs,
\eeq
where $Q_L,P_L,P_R,Q_R$ are polynomials of degree $n_L,N_L,N_R,n_R$ 
respectively and can be chosen to be with leading coefficient equal to 1
($t_0$ is a constant, possibly dimensionfull).

To use this curve as a starting point for deformations, we must identify 
explicitly its dependence on the parameters and moduli.
We will show that the curve is 
\beql{N=2-curve} G(v,t):=\Lml Q_L(v)t^3-P_L(v)t^2+P_R(v)t-\Lmr Q_R(v)=0 \eeq
where
\beql{P-def}
   P_L(v)\sce\det(v-v_L-\vev{A_L}) \hsc P_R(v)\sce\det(v-v_R+\vev{A_R}) \hs,
\eeq
\beql{Q-def}
   Q_L(v)=\det(v-v_L+m_L/\sqrt{2}) \hsc Q_R(v)=\det(v-v_R-m_R/\sqrt{2}) \hs,
\eeq
\[ v_L-v_R=m_F/\sqrt{2}  \]
and $\Lml,\Lmr$ are the instanton factors (the one-loop
  beta function coefficients being $b_{L,R}=2N_{L,R}-(N_{R,L}+n_{L,R})$). 
The definition of $P_L,P_R$ deserves some explanation. Consider, for example,
$P_L$ and take $v_L=0$ (by shifting $v$). It is a polynomial in $v$
\[ P_L(v)=\sum_{i=0}^{N_L}s_{Li}v^{N_L-i}=\prod_{a=1}^{N_L}(v-A_{La}) \hs, \]
with the coefficients $s_{Li}$ being the moduli parameterizing the space of
vacua. Alternatively, one can take the roots $\{A_{La}\}$ of $P_L$ as the
coordinates, with the understanding that they should be identified under
permutation. What eq. (\ref{P-def}) means is that {\em in the semi-classical
region} $\{A_{La}\}$ are the eigenvalues of $\vev{A_L}$ or, equivalently, that
$P_L(v)$ is the characteristic polynomial of $\vev{A_L}$. We will use this
notation also later, ``$\sce$'' meaning ``semi-classically equal''. Note that
this is an equality only up to terms vanishing in the semi-classical region.

To derive eq. (\ref{N=2-curve}), one considers first the
limits of vanishing gauge couplings. For example, a vanishing $\Lmr$ 
corresponds to the right $\NS$ brane ($S_R$) being taken to $x_6\goto\infty$
(recall that the gauge coupling $\tau$ is proportional to the length of the
finite $D_4$ branes). This is achieved by taking $C_R=0$:
\[ t[C_LQ_L(v)t^2-P_L(v)t+P_R(v)]=0 \hs. \]
the first factor corresponds to a $\NS$ brane at $t=0$ (which is indeed 
$x_6=\infty$), while the second factor should corresponds to a curve for an
$SU(N_L)$ gauge theory with $n_L+N_R$ flavors. A Change of variables
$t\goto t/C_LQ_L$ and then $y=2t-P_L$ leads to the curve
\[ y^2=P_L^2-4C_LQ_LP_R \hsc \]
which is the familiar curve for this model \cite{HO9505,APS9505}%
\footnote{Note that in these works $\lm=\sqrt{2}$ was absorbed in $m$: 
  $m/\sqrt{2}\goto m$.},
and this leads to the identification
of $C_L,P_L,Q_L$ and $P_R$, up to terms proportional to $\Lmr$.
At this stage it would be natural to choose $v_L=0$ (which also implies
$v_R=-m_F/\sqrt{2}$), but the freedom to shift
$v$ leads to a general $v_L$. The limit of vanishing $\Lml$ leads similarly
to the identification%
\footnote{To obtain the curve in the familiar form, one should choose $v_R=0$
  and change variables $v\goto-v$. We also use the freedom to redefine 
  $(-1)^{N_L+n_R}\Lmr\goto\Lmr$.}
of $C_R$ and $Q_R$.

At this stage the curve is identified up to terms proportional to $\Lml\Lmr$.
To show that there are no such terms, consider an 
$SU(\hat{N}_L)\otimes SU(\hat{N}_R)$ model, with $\hat{N}=N+1$ and choose
moduli of the form
\[ \hat{A}_{La}=\left\{\begin{array}{ll}
               A_{La}+M&a<\hat{N}_L\\-N_LM&a=\hat{N}_L\end{array}\right. \hsc
   \hat{A}_{Ra}=\left\{\begin{array}{ll}
               A_{Ra}-M&a<\hat{N}_R\\N_RM&a=\hat{N}_R\end{array}\right. \hs, \]
with a large $M$ (which will eventually be taken to infinity).
At the scale $M$ the gauge symmetry is broken to $SU(N_L)\otimes SU(N_R)$ and
$A_{La},A_{Ra}$ are the moduli of the resulting model. Assuming asymptotic
freedom ($b_L,b_R>0$), for $M$ large enough the symmetry breaks at weak
coupling and can be analyzed semi-classically. $M$ contributes, through
the moduli, to the effective mass of the fundamentals (as can be seen
by examining
the superpotential), so if the mass parameters are kept constant the effective
mass will diverge. To cancel this effect, we shift also the masses
\[ \hat{m}_L=m_L-\sqrt{2}M \hsc \hat{m}_R=m_R+\sqrt{2}M \]
(note that $M$ does not contribute to the mass of the bi-fundamentals, so
we can keep $m_F$ constant). Finally, the matching of scales is%
\footnote{This is always true at weak coupling and, therefore, is exactly
 true in the limit $M\goto\infty$, since we assume asymptotic freedom.}
\[ 
 \hat{\Lm}_L^{\hat{b}_L}=\frac{(N_LM)^2}{N_RM}\Lml=\frac{N_L^2}{N_R}M\Lml \hsc
 \hat{\Lm}_R^{\hat{b}_R}=\frac{(N_RM)^2}{N_LM}\Lmr=\frac{N_R^2}{N_L}M\Lmr \hsc
\]
where the numerators correspond to the heavy gauge bosons and the denominators
correspond to heavy fundamentals (coming from the bi-fundamentals). This 
process should be reflected correctly in the curve. The curve for the
$SU(\hat{N}_L)\otimes SU(\hat{N}_R)$ model is 
\[  \hat{\Lm}_L^{\hat{b}_L}\hat{Q}_L(\hat{v})\hat{t}^3
    -\hat{P}_L(\hat{v})\hat{t}^2+\hat{P}_R(\hat{v})\hat{t}
    -\hat{\Lm}_R^{\hat{b}_R}\hat{Q}_R(\hat{v})=0 \]
and using the scale matching, one obtains 
\beql{break}  \Lml\hat{Q}_L(\hat{v})t^3
    -\frac{1}{N_LM}\hat{P}_L(\hat{v})t^2+\frac{1}{N_RM}\hat{P}_R(\hat{v})t
    -\Lmr\hat{Q}_R(\hat{v})=0 \hsc t=\frac{N_R}{N_L}\hat{t} \hs.
\eeq
{}From eqs. (\ref{P-def},\ref{Q-def}) we deduce
\[ \hat{P}_L(\hat{v})\sce(v-v_L+N_LM)P_L(v) \hsc 
   \hat{Q}_L(\hat{v})\sce Q_L(v) \hsc \hat{v}=v+M \]
(and the same for $P_R,Q_R$), so the terms in eq. (\ref{break}) must be finite
in the limit $M\goto\infty$. These terms are functions of $\hat{v}$, $\hat{A}$,
$\hat{m}$ and $\hat{\Lm}^{\hat{b}}$, which all depend polynomially
on $M$, so if we assume that $\hat{P}$ and $\hat{Q}$ depend polynomially on
their arguments, we conclude that a term proportional to 
$\hat{\Lm}_L^{\hat{b}_L}\hat{\Lm}_R^{\hat{b}_R}$ cannot appear in them (since 
such a term would be at least quadratic in $M$).
Observe that corrections to $P$ which are linear in $\Lm^b$ are
  still allowed, so the moduli are defined only up to terms linear in
  $\Lm^b$.
At this stage we determined completely the curve for the ``up'' model, but now
one can break the symmetry, as described above, and find that the curve for
the ``down'' model also does not have $\Lml\Lmr$ terms and, therefore, is
given by eq. (\ref{N=2-curve}).

The curve (\ref{N=2-curve}) implies the identification of charges for the 
coordinates $v,t$ (in complete agreement with the type IIA analysis). In 
particular, the mass dimensions are $[v]=1$ and $[t]=N_R-N_L$.
\supnote{The two useful symmetries (see Appendix \ref{sym}) can be chosen
to be:
\nlb $R=q_\th+q_L+q_R+q_F$ which acts trivially on everything except
$\tilde{Q}Q$
and $\tilde{F}F$ and guarantees the existence of a quantum Coulomb branch
in the present case;
\nlb $q_m=\half q_\th+q_{AL}+q_{AR}$ which is the mass dimension of all the
above parameters.}

In the next subsections, we will consider the rotation of a $\NS$ brane. Such
a rotation will break $N=2$ supersymmetry and, consequently, one cannot
exclude {\em a priori}
the possibility that the Yukawa couplings $\kp$ and $\lm$
will vary. It will turn out that the curve (\ref{N=2-curve}) will continue to
play a role in these scenarios and then it will be important to reintroduce
$\kp$ and $\lm$ into the curve. This dependence is completely determined by 
symmetries (after a consistent choice of charges for the coordinates is made)
and the result is (see Appendix \ref{sym}) that one should perform the
following redefinitions
\beql{recover}
  A\goto\frac{1}{\sqrt{2}}\kp A \hsc
  \mu\goto2\mu/\kp^2 \hsc
  m\goto\kp\lm\inv m \hsc
  \Lm^b\goto\Lm^b\det(\lm/\kp)(\kp/\sqrt{2})^{2N} \hs.
\eeq
To simplify the notation, we will continue to use the form (\ref{N=2-curve})
of the curve with the understanding that each parameter in it actually 
represents an expression containing also $\kp$ and/or $\lm$ factors, as 
described in eq. (\ref{recover}).

\subsection{Rotation of a \boldmath $\NS$ Brane}
\secl{rotate}

In this subsection we explore the possibilities to rotate a $\NS$ brane, \ie,
we look for M-theory fivebranes that will correspond, in the type IIA
description, to a configuration obtained from the $N=2$ supersymmetric one 
(described in the previous subsection) by a rotation of {\em one} of the
$\NS$ branes in the ($v,w$) plane.

Consider, for example, a configuration with $S_L$ rotated by an angle $\th$
and $n_L=n_R=0$.
The corresponding quantum system will be characterized by the asymptotic
behavior of the fivebrane.
\internote{This is not a precise statement.
This may be discussed in section 5.}
In the present case, there will be three parts 
of the curve that extend to infinity in the internal space, corresponding to
the three $\NS$ branes. In each of these parts $v$ extends to infinity 
(assuming $\th\neq\half\pi$). The $w$ behavior is determined by the chosen
orientation of the $\NS$ branes and the $t$ behavior is determined by the 
bending of the $\NS$ brane, because of the $D_4$ branes attached to it
\cite{Witten9703} (see subsection \ref{IIA-quant}). This leads to the 
following behavior:
\begin{eqnarray}\nonumber
  {\mbox{\boldmath} S_L:} & t\sim v^{N_L}/C_L \hsc &
  w\sim\mu v \hsc \mu=e^{i\vph}\tan\th \hs; \\ \label{SLR-ass}
  {\mbox{\boldmath} S_M:} & t\sim v^{N_R-N_L} \hsc & w\goto w_M \hs; \\  
  \nonumber
  {\mbox{\boldmath} S_R:} & t\sim C_Rv^{-N_R} \hsc & w\goto w_R \hs,
\end{eqnarray} 
where $C_L$, $C_R$, $w_M$ and $w_R$ are parameters (the absence of a 
corresponding $C_M$ parameter in $S_M$ reflects a choice of normalization of
$t$). Their physical meaning is undetermined at this stage and will be 
identified later%
\footnote{$C_L$ and $C_R$ will, indeed, be identified with the corresponding
parameters in eq. (\ref{N=2-gen}), but this is not needed at this stage.}.
Note that classically, one would expect $w_M=w_R=0$, however we will see that
this is impossible.

We look for a Riemann surface $\Sg$, with the asymptotic behavior 
(\ref{SLR-ass})%
\footnote{We use here arguments appearing in \cite{HOO9706} for SQCD ($SU(N)$
  gauge group).}.
Each such surface will correspond to a vacuum of the
quantum system characterized by these asymptotic conditions. We first observe
that $\Sg$ can be compactified by adding to it the three points at 
$v\goto\infty$. Indeed, in each of the asymptotic regions $v$ is a local
holomorphic coordinate: $t$ and $w$ are single valued functions of $v$
(this is a reflection of the fact that each such region corresponds to a
{\em single} $\NS$ brane). This means that each $v=\infty$ point has a
neighborhood holomorphically parameterized by $1/v$ and by adding these
neighborhoods to $\Sg$ we obtain a compact Riemann surface (which will be
denoted by the same symbol $\Sg$). Next
consider the nature of $w$ as a function on $\Sg$. This is a coordinate of the
embedding of $\Sg$ in the (internal part of) spacetime and as such, it is
obviously single-valued on $\Sg$. Moreover, from the asymptotic conditions
(\ref{SLR-ass}) we see that it has a single singular point (where it diverges)
and this point is a simple pole (as can be seen, using the holomorphic 
coordinate $1/v$). If $\Sg$ is an irreducible surface (\ie, not a union of
surfaces), then this implies that $w$ is a (holomorphic) bijection between
$\Sg$ and the $w$ plane (this is why $w_M$ and $w_R$ cannot both vanish).
We therefore arrive at the following conclusions:
\begin{itemize}
\item $\Sg$ has a genus 0;
\item $w$ is a global coordinate on $\Sg$, which means that the surface can
  be described by two functions
  \beq v=V(w) \hsc t=T(w) \eeq 
  (rather then two equations).
\end{itemize}
While the second conclusion is important technically (and will be used later),
the first one is the physically significant one, because it implies that the
above rotation makes all the vector fields massive. For SQCD this is
obvious,
because the rotation corresponds to adding a mass to the adjoint,
and this lifts the Coulomb phase also classically. However, in the present
model there are also $A_R$ moduli (corresponding to the location of the right
finite $D_4$ branes in the $v$ direction) which, in the classical analysis,
are
not affected by such a rotation and parameterize a branch with massless gauge 
fields. We will see later that this branch collapses to a discrete set of
vacua.

This is a general phenomenon. Note that it arises rather directly from the
fact
that a {\em single} $\NS$ brane is being rotated. It, therefore, remains valid
also for $n_L,n_R\neq0$ and, moreover, for models with more gauge group
factors (\ie,  more $\NS$ branes). It implies that a non-zero mass $\mu$ for a
single adjoint (the first or the last in the chain) lifts the Coulomb
branch parameterized by {\em all} the adjoints. 

The above discussion refers to the case of an irreducible curve. This case 
will be further analyzed in subsection \ref{non-deg}. From the above discussion
we see that the only possibility for a Coulomb phase to survive a $\NS$
brane rotation (\ie, turning on a mass for one of the adjoints) is if the
curve is reducible. Then the rotation will not affect the whole curve (in fact,
only one irreducible component will be affected), and the unaffected
components may have non-vanishing genus and give rise to massless vector
fields in the effective 4 dimensional field theory. Such brane
configurations 	are the subject of the next subsection.

\subsection{A Reducible Curve} 
\secl{Deg}

If the curve $\Sg$ is reducible (\ie\ a union of several Riemann surfaces),
there is exactly one irreducible component containing the 
rotated $\NS$ brane and the above discussion refers to it without any changes.
For the other components, the asymptotic conditions characterizing them are
independent of $\mu$, therefore, these components are $\mu$-independent, \ie,
unaffected by the rotation. $w$ is bounded in these components and, therefore,
constant (this constant is determined by the asymptotic conditions). The 
genus of these components is not restricted by the above considerations, 
therefore, such a configuration {\em can} correspond to a Coulomb branch. The
effective gauge coupling in such a branch will be $\mu$-independent.

Assuming that the curve depends smoothly on $\mu$, if it is reducible after
rotation, it will be reducible also before rotation, so we can analyze this
kind of rotation possibilities looking at the unrotated curve, eq. 
(\ref{N=2-curve}). Consider an irreducible component $\Sg_0$ of this curve. It
may contain $l$ $\NS$ branes, where $l=0,1,2,3$. 
In the $l=0$ case, $v$ is bounded and, therefore, constant so this case
corresponds to a flat, infinite, $D_4$ brane. There is no $\NS$ to rotate 
here, so this case is irrelevant for the present discussion%
\footnote{Recalling that
  the semi infinite $D_4$ branes can be seen as ending on $D_6$ branes at
  infinity, it is clear that this is a root of a Higgs (or mixed) branch where
  a $D_4$ brane extends between two $D_6$ branes and can ``slide'' in the (789)
  directions.}.
The $l=3$ case is essentially the same as the irreducible case (that will be
discussed later), with reduced $n_L,N_L,N_R$ and $n_R$ (this is the complement
of the $l=0$ case).
The $l=2$ case is mathematically identical to the description of an $SU(N)$
model and the rotation of one of the two $\NS$ branes was described in 
\cite{HOO9706,Witten9706,BIKSY9706}. We will encounter such a situation as a
special case in subsection \ref{non-deg}

Finally, consider the $l=1$ case, in which the rotated $\NS$ brane is 
detached from the remaining two. The component to be rotated has one region
of diverging $v$ and there we have (according to the asymptotic conditions)
$t\sim Cv^k$ for some (integral) $k$. Therefore, $v$ is a global holomorphic
coordinate and the curve is described by a holomorphic function $t=T_0(v)$.
If this function is bounded, then it is constant, implying also $k=0$, and
this corresponds to the detachment of a {\em flat} $\NS$ brane. For
$n_L=n_R=0$
this is the only possibility. For non-vanishing $n_L$ and/or $n_R$, there are
more possibilities, corresponding to a $\NS$ brane connected to semi-infinite 
$D_4$ branes. In the following, we concentrate on the detachment of flat
$\NS$ branes.

\subsubsection{A Flat \boldmath $\NS$ Brane}
\secl{Flat}

The simplest possibility for the curve to be reducible is the case of a flat
$\NS$ brane detaching from the rest of the curve. This corresponds to $G$
in eq. (\ref{N=2-curve}) being divisible, as a polynomial, by $(t-t_0)$ for 
some ($v$-independent) $t_0$. Equivalently, the equation
\beql{v-ident} \Lml Q_L(v)t_0^3-P_L(v)t_0^2+P_R(v)t_0-\Lmr Q_R(v)=0 \eeq
should be an identity in $v$. Comparing the leading powers of $v$, we conclude
that at least two of the 4 polynomials in eq. (\ref{v-ident}) should have the
highest degree. Assuming $b_L,b_R>0$ (asymptotic freedom), we obtain two
different possibilities:
\begin{itemize}
\item $n_L<N_L=N_R>n_R$

  Comparing powers of $v$, one obtains the conditions
  \beql{SM-Cond} P_L-\Lml Q_L=P_R-\Lmr Q_R=:\bar{P} \eeq
  and when they are satisfied, $G$ becomes
  \beql{SM-curve} G=(t-1)(\Lml Q_Lt^2-\bar{P}t+\Lmr Q_R) \eeq
  which corresponds to the detachment of the {\em central} $\NS$ brane.
  The condition (\ref{SM-Cond}) for the detachment implies that $P_L,P_R$
  (and $\bar{P}$) must coincide in the classical limit (and, in particular,
  $m_F=0$), which means that classically the finite $D_4$ branes from both
  sides of $S_M$ must be aligned with each other. This is exactly the
  criterion obtained from a ``classical'' analysis, in both the geometric
  (Type IIA brane) and algebraic (field-theoretical)
  descriptions%
\footnote{This is the type 1 situation of section \ref{FT} and $P_L=P_R$
    corresponds to $A_L=-A_R$ and $m_F=0$.}.
  The quantum corrections to this criterion are, in most cases, shifts of the
  moduli, which are insignificant in the present context, since we
  characterized them only classically%
\footnote{This will become significant in the investigation of the connections
  to other models and branches, where the moduli have a more unambiguous
  definition.}.
  For $n_L=N_L-1$ and/or $n_R=N_R-1$ there is also a shift in the condition
  on $m_F$:
  \beql{mF-mod} 
    \frac{1}{\sqrt{2}}m_F\equiv v_L-v_R=\frac{1}{N}(\ep_R\Lm_R-\ep_L\Lm_L)
  \eeq
  (where $\ep=1$ when $n=N-1$ and vanishes otherwise). 
  We interpret this shift as follows: classically, a detachment situation is
  a root of a branch parameterized by $\det\vev{F}$ and $\det\vev{\tilde{F}}$,
  therefore, one
  expects that the (exact quantum) mass of $F$ should vanish there (indeed, at
  this point the curve degenerates, reflecting the existence of massless 
  states). This
  suggests that the above value for $m_F$ is the classical
  value needed to cancel the
  quantum corrections to the mass, so that the final mass will vanish.

\item $n_L=N_L>N_R>n_R$

  Comparing powers of $v$, one obtains the conditions
  \beql{SL-Cond} P_L-Q_L=\Lml(P_R-\Lml\Lmr Q_R)=:\Lml\bar{P}_R \eeq
  and when they are satisfied , $G$ becomes
  \beq G=(\Lml t-1)(Q_Lt^2-\bar{P}_Rt+\Lmr Q_R) \eeq
  which corresponds to the detachment of the {\em left} $\NS$ brane.
  The condition (\ref{SL-Cond}) for the detachment implies that $P_L$ and
  $Q_L$ must coincide in the classical limit, which means, as in the previous
  case, that classically the $D_4$ branes from both sides of the detaching
  $\NS$ brane must be aligned with one another. $\bar{P}_R$ coincides with
  $P_R$ classically, and the quantum difference is only in the moduli%
\footnote{Observe that in this case $b_R>0$ implies $N_R-n_R\ge2$, so the 
  sub-leading terms of $\bar{P}_R$ and  $P_R$ coincide.}.

\end{itemize}

The detachment of the right $\NS$ brane is treated analogously.

\subsubsection{Rotating a Flat Central \boldmath $\NS$ Brane} 
\secl{rot-flat}

We consider now the rotation of a (detached) flat central $\NS$ brane, which
is possible, as we saw, only for $N_L=N_R=:N$.
According to the identification in subsection \ref{Identify}, such a rotation
corresponds, classically, to $\mu_L=-\mu_R=:-\sqrt{2}\mu\neq0$ and $m_F=0$.
As discussed in subsection \ref{FT-quant}, at scales below $\mu$, $A_L$ and
$A_R$ decouple, leading to a model with a tree level superpotential%
\footnote{This differs from eq. (\ref{Weff}) because here $\kp$ and $\lm$ are
  set to $\sqrt{2}$ and not 1.}
\beql{Weff-m}
  W = \tr(m_L\tilde{Q}_LQ_L)+\tr(m_R\tilde{Q}_RQ_R)
    +\frac{\sqrt{2}}{\mu}[\tr(\tilde{Q}_LBQ_L)-\tr(\tilde{Q}_RBQ_R)] \hs,
\eeq
with
\[ B:=F\tilde{F}-\frac{1}{N}I_N\tr(F\tilde{F}) \]
(and $A_L\sce-A_R\sce B/\mu$),
so the resulting effective model, without adjoint fields, will have
the tree-level superpotential (\ref{Weff-m}). This model has vacua
parameterized
by $F$ and $\tilde{F}$ and we have shown that they are not lifted by quantum
corrections. 

The curve (the unrotated part; see eq. (\ref{SM-curve})) becomes
\beql{SMR-curve} \Lmlp Q'_Lt^2-\bar{P}'t+\Lmrp Q'_R=0 \hsc \eeq
where%
\footnote{$m_R$ may be corrected quantum mechanically: $m_R\goto m_R-m_F$, 
  where $m_F$ is given by eq. (\ref{mF-mod}).}

\[ w=\mu v \hsc 
   \bar{P}'(w):=\mu^N\bar{P}\sce\mu^N\det(v-A_L)\sce\det(w-B) \hs, \]
\[ Q'_L(w)=Q_L(v)=\det(w/\mu+m_L/\sqrt{2}) \hsc
   Q'_R(w)=Q_R(v)=\det(w/\mu-m_R/\sqrt{2}) \]
and 
\beql{match-mu} \Lmlp=\mu^N\Lml \hsc \Lmrp=\mu^N\Lmr   \eeq
are the usual matching relations for the scales. The geometric interpretation
of eq. (\ref{SMR-curve}) is as follows: the relation $w/v=\mu$ identifies $w$
as the coordinate of the $(89)$ plane (according to the classical analysis, 
which is valid in the semi-classical -- asymptotic -- region of the curve),
so actually $S_L$ remains in the original orientation ($v$ plane) and the rest
of the curve is rotated with respect to it. Observe that for fixed $A$, $m$
and $\Lm^b$, the projection of the curve on the $v$ plane (which is what eq.
(\ref{SM-curve}) describes) remains unchanged. One consequence of this is that
the effect of non-vanishing $\mu$ can be undone by a holomorphic change of 
coordinates $w-\mu v\goto v$, so the low-energy effective gauge coupling is
independent of $\mu$ (for given $A$, $m$ and $\Lm^b$). Another consequence is
that strictly speaking, the effect of changing $\mu$ is not a {\em rotation}
but, rather a {\em stretching} of the curve in the $w$ direction.
{}From this it is also clear that taking $\mu$ to infinity holding $A$, $m$ and
$\Lm^b$ fixed will not lead to a curve in the $w$ plane. This is the
geometrical origin of the change of scales (\ref{match-mu}). Holding
${\Lm'}^{b'}$ (and $B$) fixed, corresponds, roughly
to stretching the curve in the $v$ direction and this is the way to obtain
a finite curve for $\mu\goto\infty$. For $n_L=n_R=0$, this is an exact
statement%
\footnote{When there are fundamentals,
  their mass parameters $m_L,m_R$ determine the asymptotic $v$ coordinate of
  the semi-infinite $D_4$ branes, so if they are kept constant when
  changing $\mu$, then the asymptotic $w$ coordinate changes and in the limit
  $\mu\goto\infty$ these $D_4$ branes escape to infinity.
  To fix this, one should recover $\lm$ (using (\ref{recover})) and rescale it,
  holding $\lm/\mu$ fixed.},
so the projection of the curve on the $w$ axis is independent of $\mu$, for
given ${\Lm'}^{b'}$ and $B$ and, therefore, so is the effective gauge
coupling (as can be verified by a shift $v-w/\mu\goto v$).

For $n_L=n_R=0$ we obtain (at scales below $\mu$), an $N=1$
supersymmetric $SU(N)^2$ gauge group,
with two bi-fundamentals and a vanishing superpotential. This 
model was analyzed in \cite{CEFS} by purely field theoretical methods. The
analysis led to a curve that is equivalent with (\ref{SMR-curve}) (with a
modified definition of moduli). A more general family of models, corresponding
to $n_L\neq0=n_R$ and a superpotential 
\[ W=\sum_l\tr[h_l\tilde{Q}_L(F\tilde{F})^lQ_L] \]
was analyzed in \cite{Gremm9707}. When $h_l$ vanishes for $l\neq0$, these
models coincide with the present models (with $h_0=m_L$) and, again, so do the
curves.
\supnote{Actually, in \cite{Gremm9707} the effect of $\tr F\tilde{F}$ is
  ignored, so it is plausible that the results of that work are actually for
  $W=\sum_l\tr[h_l\tilde{Q}_LB^lQ_L]$. Indeed, when $h_l$ vanishes for
  $l\neq0,1$, the curves are as obtained here.}
The curves for $n_Ln_R\neq0$ and/or non vanishing Yukawa coupling (finite
$\mu$) obtained here, seem to be unknown so far.

Consider now a large vev for $F$
\[ \vev{F}=cI_N \hsc c\gg\Lm_L,\Lm_R \]
(note, however, that we still assume that $\mu$ is larger than anything else
so, in particular, $\mu\gg c$). In gauge invariant coordinates, this means a
large $D:=\det F=c^N$. At the scale $c$ the gauge symmetry is broken 
to $SU(N)_D$ and if $c\gg\Lm_L,\Lm_R$, this can be analyzed semi-classically.
At scales below $c$ we have, effectively, an $SU(N)_D$ model with an adjoint
$A_D=\tilde{F}-\rec{N}\tr\tilde{F}$, two singlets $\tr F,\tr\tilde{F}$ and
fundamentals. The corresponding curve, as obtained from (\ref{SMR-curve}) is
\beql{SMD-curve} t_D^2-P_Dt+\Lmd Q_D=0 \hsc \eeq
where
\[ x=w/c \hsc t_D=t/D \]
\[ P_D(x):=\bar{P}'/D\sce\det(w-B)/D\sce\det(x-A_D) \hs, \]
\[ \Lmd=\Lmlp\Lmrp/D^2 \hsc Q_D(v_D)=Q_LQ_R|_{v=\frac{c}{\mu}x} \hs. \]
For $\mu\goto\infty$, $Q_D\goto\det m_L\det m_R$ and we recover the known
curve for $SU(N)$ with an adjoint, fundamentals and vanishing Yukawa
coupling \cite{HO9505}.

\supnote{
  $\bullet$ GEOMETRICAL MEANING: $w$ is determined by $B\sim\tilde{F}F$,
  which diverges, while $x\sim w/c$ remains finite; not illuminating.
  \nl PROBLEMS WITH FINITE $\mu$:
  \nlb We have dependence on $c$, which is not gauge invariant. What does it
  mean?
  \nlb Can we have $c>\mu$? presumably, in this case the integration out of 
  $A$ is not justified, and one should return from ${\Lm'}^{b'}$ to $\Lm^b$}

\subsection{An Irreducible Curve}
\secl{non-deg}

In this subsection we return to the consideration of a rotation of a $\NS$
brane when the curve is irreducible. We restrict ourselves to $n_L=n_R=0$
and consider the rotation of the left $\NS$ brane. As shown at the beginning
of this section, the corresponding curve can be described by two functions 
\beql{VT} v=V(w) \hsc t=T(w) \eeq
and they are characterized by the asymptotic conditions (\ref{SLR-ass}). 
For $V(w)$ the conditions are that it has two simple%
\footnote {The poles are simple because, as discussed above, $v$ is a local
  holomorphic coordinate in the neighborhood of each pole.}
poles at $w=w_M$ and at $w=w_R$ while $V\sim w/\mu$ for $w=\infty$.
This leads to the general form 
\beql{V-func} 
  V=\frac{p(w)}{\mu(w-w_M)(w-w_R)} \hsc p(w)=w^3+p_1w^2+p_2w+p_3 \hs,
\eeq
where the polynomial $p(w)$ does not vanish at $w_M$ and $w_R$.

Now the conditions for $T(w)$ imply, up to multiplicative constants
\begin{eqnarray}\nonumber
  {\mbox{\boldmath} S_L:} & t\sim w^{N_L} & \hsc
  \mbox{for }w\goto\infty \hs; \\ 
  {\mbox{\boldmath} S_M:} & t\sim (w-w_M)^{N_L-N_R} & \hsc 
  \mbox{for }w\goto w_M \hs;\\
  \nonumber
  {\mbox{\boldmath} S_R:} & t\sim (w-w_R)^{N_R} & \hsc
  \mbox{for }w\goto w_R \hs
\end{eqnarray} 
and this leads to the general form 
\beql{T-func} T=C(w-w_M)^{N_L-N_R}(w-w_R)^{N_R} \hsc C\neq0 \hs. \eeq
Substituting formulas (\ref{V-func}) and (\ref{T-func}) into the asymptotic
conditions (\ref{SLR-ass}) leads to the following relations:
\begin{eqnarray}\nonumber
  {\mbox{\boldmath} S_L:} && C=(C_L\mu^{N_L})\inv \hs; \\ 
  \label{SLR-rel} 
  {\mbox{\boldmath} S_M:} && 
  p(w_M)^{N_L-N_R}=\frac{C_L\mu^{b_L}}{(w_M-w_R)^{b_R}} \hs; \\  
  \nonumber
  {\mbox{\boldmath} S_R:} &&
  p(w_R)^{N_R}=C_LC_R\mu^{N_L+N_R}(w_R-w_M)^{b_R} \hs.
\end{eqnarray}
Given $p(w_M)$ and $p(w_R)$, $V(w)$ is essentially determined%
\footnote{There is one free parameter left in $p(w)$, which corresponds to the
  arbitrariness in the choice of origin for $v$.},
therefore, we have the following situation:
\begin{description}
\item{\boldmath $N_L\neq N_R$ \bf:} The asymptotic behavior defines
  (for any $w_M\neq w_R$!) $N_R|N_L-N_R|$ solutions;
\item{\boldmath $N_L=N_R=:N$ \bf:} $w_M$ and $w_R$ are restricted by
  $(w_M-w_R)^N=C_L\mu^N$ and when this is satisfied, there are $N$ one 
  (complex) parameter families of solutions.
\end{description}
To obtain more information about the curve, we use 
the following observation \cite{HOO9706}. Let $\Sg_0$ 
be the projection of the curve $\Sg$ on the $(t,v)$ subspace. We expect that
$\Sg$ depends continuously on $\mu$ in the neighborhood of $\mu=0$
and this leads also to the expectation that $w$ is a single valued function
on $\Sg_0$ (since it is the constant function for $\mu=0$). This means that
$\Sg$ has the following description:
\[ G(v,t)=0 \hsc w=W(v,t) \]
(where $G=0$ defines $\Sg_0$). At this point we have to specify more
precisely what we are looking for: we look for a {\em one parameter} family
(parameterized by $\mu$) of deformations of the curve (\ref{N=2-curve}),
which means that $\mu$ is the only additional parameter determining $\Sg$
beyond those in (\ref{N=2-curve}).
\supnote{This is a delicate point, because it is not clear how to exclude
  the appearance of $F$ moduli. In fact, it is possible that they appear, and
  determine $w_M-w_R$.
  Such moduli will carry a $U(1)_{89}$ charge, so a combination
  like $\tilde{F}F/\mu$ will be neutral and, therefore, not excluded by the
  considerations below. To avoid such complications, we set these moduli, if
  they appear, to zero.
  \nl Perhaps one can show $\mu$ and $F$ independence, so it is not really
  necessary to set the $F$ moduli to 0. In this context it will be useful
  to check if by avoiding the assumption of a one parameter family one obtains
  more solutions.}
{}From table (\ref{U1RT}) we see that $\mu$ and $w$ are the only quantities in
$\Sg$ that carry a $U(1)_{89}$ charge.
Now, since $G$ is independent of $w$, $U(1)_{89}$ symmetry implies that it
is also independent of $\mu$ and continuity in $\mu$ implies that $\Sg_0$
is exactly the curve for $\mu=0$, \ie, $G$ is given by eq. (\ref{N=2-curve})%
\footnote{Such a phenomenon was already observed explicitly in subsection
  \ref{rot-flat}.}.
One consequence of this is the identification
\[ C_L=\Lml \hsc C_R=\Lmr  \] 
(as can be seen by comparing the asymptotic conditions to the corresponding
behavior of the curve). But much more information comes from the fact that
\[ G(v,t)|_{v=V,t=T}=0 \]
(where $V$ and $T$ are taken from eqs. (\ref{V-func},\ref{T-func})) is an
identity in $w$. After the substitution $v=V,t=T$, this identity takes the
following form:
\supnote{One can take $\mu=1$ and recover $\mu$ later by $U(1)_{89}$ symmetry:
  \nl $w\goto w/\mu$, $C\goto\mu^{N_L}C$.}  
\begin{eqnarray}\nonumber
 0&=& [C_Lt^3-P_Lt^2+P_Rt-C_R]_{v=V,t=T}(w-w_M)^{b_R}\mu^{N_R} \\ \label{ident}
  &=& \mu^{N_R}C_LC^3(w-w_M)^{b_L}(w-w_R)^{3N_R} \\ \nonumber
  & & -\mu^{N_R-N_L}C^2(w-w_R)^{b_R}
      \sum_{k=0}^{N_L}l_k(w-w_M)^k(w-w_R)^kp(w)^{N_L-k} \\ \nonumber
  & & +C\sum_{k=0}^{N_R}r_k(w-w_M)^k(w-w_R)^kp(w)^{N_R-k}
      -\mu^{N_R}C_R(w-w_M)^{b_R} \hs,
\end{eqnarray}
where $l_k,r_k$ are the coefficients in $P_L,P_R$ respectively:
\beql{lr-derf}
  P_L(v)=\sum_0^{N_L}l_kv^{N_L-k} \hsc 
  P_R(v)=\sum_0^{N_R}r_kv^{N_R-k}
\eeq
(note that $l_0=r_0=1$ while $l_1=-N_Lv_L$ and $r_1=-N_Rv_R$).
Expanding in powers of $w$, one obtains a set of equations for the parameters
in eqs. (\ref{V-func}), (\ref{T-func}) and (\ref{lr-derf}). From
now on we assume, for simplicity, $b_L,b_R\ge2$ (otherwise there are
additional terms in the expressions given below). Furthermore, we choose
$w_R=0$ (by a shift of $w$). There are 4 equations that do not
depend on the moduli $l_k,r_k$, $k>1$. These are the coefficients of 
$w^{2(N_L+N_R)},w^{2(N_L+N_R)-1},w^1$ and $w^0$ and they lead, correspondingly,
to
\begin{eqnarray}
\label{C-solve} C & = & (\mu^{N_L}\Lml)\inv \hs, \\
\label{p1-solve} p_1 & = & \mu v_L-\frac{b_L}{N_L}w_M \hs, \\
\label{p2-solve} p_2 & = & -\mu v_Rw_M-\frac{b_R}{N_R}\frac{p_3}{w_M} \hs, \\
\label{p3-solve} p_3^{N_R} & = & \mu^{N_L+N_R}\Lml\Lmr(-w_M)^{b_R} \hs,
\end{eqnarray}
which means
\beql{V-solve}
  V-v_L=\frac{w^3-\frac{b_L}{N_L}w_Mw^2+\mu(v_L-v_R)w_Mw
              +\left(1-\frac{b_R}{N_R}\frac{w}{w_M}\right)p_3}{
              \mu(w-w_M)w} \hs.
\eeq
We could also expand around $w_M$ (\ie, set $w_M=0$), obtaining other 4
moduli-independent equations. Only one of them (the coefficient of $(w-w_M)^0$)
contains new information (beyond eqs. (\ref{C-solve}-\ref{p3-solve})).
Translating back to the choice $w_R=0$, one obtains
\beql{wM-rel}
  p(w_M)^{N_L-N_R}=\frac{(\mu\Lm_L)^{b_L}}{(w_M)^{b_R}} \hs.
\eeq
The implications of this relation depend on whether $N_L=N_R$ or not:
\begin{description}
\item{\boldmath $N_L\neq N_R$ \bf :} Eqs. (\ref{p1-solve},\ref{p2-solve})
  imply
  \beql{pwm} p(w_M)=\mu(v_L-v_R)w_M^2+(N_L-N_R)
     \left(\frac{p_3}{N_R}-\frac{w_M^3}{N_L}\right) \hs. \eeq
  Combining this with eqs. (\ref{p3-solve}) and (\ref{wM-rel}) leads to
  \begin{eqnarray}\label{mF-solve}
    m_F/\sqrt{2} & \equiv & v_L-v_R \\ \nonumber & = &
    \left(\frac{\mu^N_L\Lml}{w_M^{N_L}}\right)^{\frac{1}{N_L-N_R}}
    +(N_L-N_R)\left[\frac{w_M}{\mu N_L}-\frac{1}{N_R}
    \left(\frac{\mu^N_L\Lml\Lmr}{(-w_M)^{N_L}}\right)^{\frac{1}{N_R}}\right]
    \hs,
  \end{eqnarray}
  so $w_M$ (or, more generally, $w_M-w_R$) can be seen as representing $m_F$
  in the asymptotic conditions (\ref{SLR-ass}).
\item{\boldmath $N_L=N_R=:N$ \bf :} Eq. (\ref{wM-rel}) fixes $w_M$:
  \beql{wM-solve} w_M^N=(\mu\Lm_L)^N \eeq
  and $m_F$ remains free. In other words, in this case the parameter $m_F$ is
  not reflected in the asymptotic conditions.
\end{description}
The other equations following from the identity (\ref{ident}) determine
(uniquely) the moduli $l_k,r_k$ in the $N=2$ curve,
\ie, the points in the moduli space
of the $N=2$ configurations from which a rotation is possible. Note that in
general there are much more equations then
variables, so one may expect that $m_F$ is also constrained or even that
there are no solutions at all. This has been checked for some non-trivial
cases and it was found that $m_F$ remains free. Presumably, this is true
in general. To clarify what exactly is left unsettled,
we return to the question of the conditions
for a rotation. So far we have shown that the vanishing genus of $\Sg$
(which translates to the complete degeneration of $\Sg_0$)
is a necessary condition
for a rotation. We now argue that it is also sufficient%
\footnote{We are grateful to Yaron Oz for helping to clarify this point.}.
Indeed, the functions $V(w),T(w)$
define a {\em normalization} of $\Sg_0$, \ie, a map
from a Riemann surface to $\Sg_0$, which is bi-holomorphic everywhere, except
at (the inverse image of) singular points of $\Sg_0$.
Such a normalization exists for any algebraic curve in $\CC^2$ \cite{norm}.
Apparently, we constrain this normalization by the asymptotic
conditions (\ref{SLR-ass}), but in fact they are always satisfied: the 
$t\leftrightarrow v$ relations follow from the asymptotic behavior of $\Sg_0$
(derivable from eq. (\ref{N=2-curve})) and the $w\leftrightarrow v$ relation
only means that the normalization is to a genus 0 surfaces, which was already
shown to be necessary. Therefore, {\em every curve in the family
(\ref{N=2-curve}) is rotatable iff it is completely degenerate.}
What remains unknown is if such curves exist for any given value of $m_F$.
\supnote{Count of solutions:
 \nl $N_L=N_R:=N$:
 \nl $w_M$ (N values); $m_F$ free $\goto$ $p_1,p_2$ (unique), $p_3$ (N values)
 \nl $N_L\neq N_R$:
 \nl $w_M$ free $\goto$ $p_3$ ($N_R$ values) $\goto$ $m_F$ ($|N_L-N_R|$ values)
 $\goto$ $p_1,p_2$ (unique).
 \nl The distinction between parameters and moduli should be clarified.}

To summarize, by the identification of the projection of $\Sg$ on the $(v,t)$
subspace, we recovered correctly
(see eqs. (\ref{C-solve},\ref{p3-solve},\ref{wM-rel})) the relations
(\ref{SLR-rel}) obtained from the asymptotic conditions (\ref{SLR-ass}).
We obtained, however, additional equations: eqs. (\ref{p1-solve}) and
(\ref{p2-solve}) that determine $p_1$
and $p_2$ (as functions of $m_F$) up to the freedom to shift $v$
and, for $N_L\neq N_R$, also correlate between $w_M$ and $m_F$; and
other equations that determine the moduli $l_k,r_k$. As a result, we obtain,
for any value of $m_F$, (at most) finite number of discrete solutions (vacua).

One can identify the situation when the central $\NS$ brane ($S_M$) detaches
from the rest of the curve. As discussed at the beginning of subsection
\ref{Deg}, this is possible only for
$N_L=N_R=:N$, and then the detached $S_M$ is flat: constant $t$ and $w$
(since in the present model there are no semi-infinite $D_4$ branes).
In the present 
description this happens whenever $p(w_M)=0$, and by eqs. (\ref{pwm}) and
(\ref{wM-solve}) this means $v_L=v_R$ (as for the $\mu=0$ case, discussed
in subsection \ref{Flat}). Setting $v_L=v_R=0$, one obtains
\[ p(w)=(w-w_M)(w^2-\frac{p_3}{w_M}) \hs, \]
leading to a degenerate curve:
\[ (w-w_M)(w^2-\mu vw-\frac{p_3}{w_M})=0 \hsc
   t=\left(\frac{w}{\mu\Lm_L}\right)^N \hs. \]
The first factor corresponds to the flat $S_M$: $w=w_M,t=1$. Substituting
\beql{wMp3}
  w_M=\ep'\mu\Lm_L \hsc p_3=-\ep\ep'\mu^3\Lm_L^2\Lm_R \hsc \ep^N={\ep'}^N=1
\eeq
in the second factor (using eqs. (\ref{wM-solve},\ref{p3-solve})), one obtains
\[ wv=\ep\mu\Lm_L\Lm_R+w^2/\mu \hsc
   t=\left(\frac{w}{\mu\Lm_L}\right)^N \hs, \]
which is identical to the curve of $N=1$ SYM with a massive adjoint
\cite{HOO9706}, where
$\mu$ is the mass of the adjoint and $\Lm=\sqrt{\Lm_L\Lm_R}$ is the scale. 
\internote{EXPLANATION? MOVING/ROTATING $S_M$}

Next we consider the limit $\mu\goto\infty$. We have already seen that if the
parameters of the initial $N=2$ curve are all held fixed, the projection of
the curve on the $(v,t)$ subspace remains unchanged, so $\mu$ does not
parameterize a {\em rotation} but rather a {\em stretching} in the $w$
direction. From this it is clear that the $\mu\goto\infty$ limit of such a
procedure will be singular%
\footnote{Such a situation was already discussed in subsection
  \ref{rot-flat}.}.
The correspondence to field theory, where $\mu$ is
identified as the mass $\mu_L$ of $A_L$, leads to the resolution of this
problem: in this process $A_L$ is decoupled, so we should hold fixed the
combination $\Lmlp=\mu^N_L\Lml$ (the instanton factor of the model without
$A_L$) and not $\Lml$. When $N_L=N_R:=N$, one obtains (see eq.
(\ref{wM-solve}))
\beq w_M^N={\Lm'_L}^{2N} \eeq
and when $N_L\neq N_R$, eq. (\ref{mF-solve}) becomes
\beql{vLvR}  v_L-v_R=
    \left(\frac{\Lmlp}{w_M^{N_L}}\right)^{\frac{1}{N_L-N_R}}
    +(N_L-N_R)\left[\frac{w_M}{\mu N_L}-\frac{1}{N_R}
    \left(\frac{\Lmlp\Lmr}{(-w_M)^{N_L}}\right)^{\frac{1}{N_R}}\right] \hs,
\eeq
so in both cases $w_M$ and $v_L-v_R$ are finite in the limit $\mu\goto\infty$.
The curve takes the form
\begin{eqnarray}
  V-v_L&=&\frac{w^3-\frac{b_L}{N_L}w_Mw^2+\mu(v_L-v_R)w_Mw
              +\left(\frac{b_R}{N_R}w-w_M\right)\mu q}{\mu(w-w_M)w} \hs, \\
  \nonumber T&=&(w-w_M)^{N_L-N_R}w^{N_R}/\Lmlp \hs,
\end{eqnarray}
where
\[ q:=-\frac{p_3}{\mu w_M}
     =\left[\Lmlp\Lmr(-w_M)^{N_R-N_L}\right]^{\frac{1}{N_R}} \hs, \]
so in the $\mu\goto\infty$ limit, $T$ remains unchanged, while $V$ becomes
\beq V=v_L+\frac{(v_L-v_R)w_M+(1-\frac{N_L}{N_R})q}{w-w_M}+\frac{q}{w} \hs.
\eeq
Each of the three terms corresponds, roughly, to the tree $\NS$ branes.
For $N_L=N_R$ one can use the explicit expressions (\ref{wMp3}), to obtain
\[ V=v_L+\frac{\ep'{\Lm'_L}^2m_F/\sqrt{2}}{w-\ep'{\Lm'_L}^2}
        +\frac{\ep{\Lm'_L}^2\Lm_R}{w} \hsc
   T=\left(\frac{w}{{\Lm'_L}^2}\right)^N \hs. \]

It is interesting to compare the above curves to those corresponding to
$N=1$ SQCD with $N_L$ colors and $N_R$ flavors.
For a uniform quark mass $m$, the curve is 
\cite{HOO9706} (in an appropriate choice of coordinates)
\beq V=v_L-\frac{w_-m}{w+w_-} \hsc
     T=(w+w_-)^{N_L-N_R}w^{N_R}/\Lmlp \hs,
\eeq
where
\[ w_-=\left(m^{N_R-N_L}{\Lm'}^{b'}\right)^{\frac{1}{N_L}} \hs. \]
Identifying $\Lm'=\Lm'_L,m=v_L-v_R,w_-=-w_M$, the curves coincide for $q=0$.
For finite $q$, the main difference between the curves is the absence (in the
SQCD curve) of the third term of $V$ -- the term that corresponds to the right
$\NS$ brane ($S_R$). The absence of this term is understood as follows:
in the type IIA brane description, the difference between the configurations
is that $S_R$ is replaced by $N$ $D_6$ branes (each connected
to one right $D_4$ brane). The third term in $V$ describes the extension of
$S_R$ in the $v$ direction and its absence in the second curve corresponds to
the fact that the $D_6$ branes are at fixed $v$. Now, the limit $q\goto0$
(for fixed $w_M$ and $\Lm'_L$) is the limit $\Lm_R\goto0$ which, in the brane
description, corresponds to taking $S_R$ to $x^6\goto\infty$. The coincidence
of the curves means that in this limit the right $D_4$ branes ``forget'' what
brane they are attached to. Another significant point in this comparison is
that the quark mass is uniform. This means that all the $D_6$ branes are in
the same $v$ positions.
Classically, this implies that all the right $D_4$ branes coincide and in our
model this means $\vev{A_R}=0$. This is in agreement with the fact that the
$N=2$ curve we rotated has genus 0, since this situation occurs for
$\vev{A_R}$ of the order of $\Lm_R$. Finally,
in SQCD, in the limit $m\goto0$ there are vacua only for $N_L\le N_R$, so it is
interesting to see what is the corresponding behavior in our model. Setting
$v_L=v_R$ (and $\mu=\infty$) in eq. (\ref{vLvR}), one obtains
$w_M\sim\Lm_R^{N_R-N_L}$, which indeed diverges in the limit $\Lm_R\goto0$,
unless $N_L\le N_R$.

\newsection{Discussion}
\secl{disc}

In the previous sections we analyzed the same field theoretical model
using: a direct (field theory) approach,
weakly coupled type IIA string theory
and an M theory limit corresponding to a strongly coupled type IIA string. 
In this
section we discuss some conclusions that can be drawn from the results.

\subsection{The Frozen $U(1)$ Factors}

The freezing of the $U(1)$ gauge factors, described in subsection 
\ref{bend}, raises a puzzle%
\footnote{This problem in not specific to the present model and appears already
in SQCD.}.
As we saw
in the comparison of the two classical models, the freezing of the $U(1)$
vector multiplets has an opposite effect on other quantities: $\xi$ and $\eta$
effectively change from being parameters to being moduli
(related to vev's of $F$ and $\tilde{F}$). Classically, we
identified these parameters as the relative location of the $\NS$ branes in the
(789) directions. 
Keeping this identification seems to conflict with the ``principle'' that
moduli should not correspond to a motion of infinite objects (this was, after
all, the principle that led to the identification of this effect in the first
place!).
So one may deduce from this principle that this identification is
wrong quantum mechanically, but this approach raises other problems.
The FI parameters are left with no geometrical
interpretation, which is acceptable (we already have such parameters -- the
Yukawa couplings of the bi-fundamentals).
On the other hand, the (789) location of the $\NS$ branes
is left with no imprint in the effective gauge theory, implying that it is
either irrelevant (which would be surprising) or fixed by quantum effects.

We may look for guidance in the results we obtained in M theory.
In subsection \ref{Flat}, where we considered parallel $\NS$ branes
(corresponding to $\mu_L=\mu_R=0$),
we found detachment of a (flat) brane exactly
in situations where we expected it from classical considerations. In
particular, $S_M$ was detached exactly where one would expect the root of the
Higgs branch, parameterized by $D\sce\det\vev{F}$ and 
$\tilde{D}\sce\det\vev{\tilde{F}}$ (compare to subsection \ref{mu0})%
\footnote{Similarly, one can easily recognize the detachment of $S_L$ as the
root of the baryonic branch parameterized by $Q_L$ moduli, although we did not
consider these branches in this work.}.
Once a
(flat) $\NS$ brane is detached, it is obviously free to move in the (789)
directions. Moreover, once the detached brane is moved, the Coulomb branch is
truncated, so this motion cannot be completely irrelevant. It does not
influence the effective $U(1)$ gauge coupling, but neither are Higgs moduli,
at least as long as the $N=2$ supersymmetry is unbroken \cite{APS9603}. In
subsection \ref{non-deg}, where we considered $\mu_L\neq0=\mu_R$, we observed
a detachment of $S_M$ for $N_L=N_R$, again, exactly where one expects
the root of the Higgs branch (compare with subsection \ref{mu-yn})%
\footnote{In the classical analysis we found a (smaller) Higgs phase also for
$N_L>N_R$. We do not see this branch in the M theory analysis so, presumably,
it is lifted by quantum corrections.}.

So the results from M theory seem to suggest that the (789)
location of $S_M$ realizes, also
in the quantum situation, the FI parameters, which are related (together
with another, non geometrical, degree of freedom) to $D$ and $\tilde{D}$.
This is also supported by symmetries and quantitative relations.
But if this is true, we are back to the original problem: how can a modulus
correspond to infinite motion. In principle, this problem can be resolved in
two ways: either the above moduli are some how frozen, or the $S_M$ brane
is not infinite. In fact, we already know a situation with the same apparent
problem. When the fundamental fields are realized by $D_6$ branes, there is a
branch of vacua, (parameterized by $Q$ moduli) corresponding to the (789)
locations of a $D_4$ branes
extended between left $D_6$ branes and right $D_6$ branes. When these $D_6$
branes are taken to $-\infty$ and $+\infty$ respectively, the $Q$-moduli seem
to correspond to the motion of infinite $D_4$ branes! Perhaps the $\NS$ branes
can also be made finite in a similar fashion. This deserves a further
investigation.

\subsection{Lifting the Coulomb Branch by Adjoint Masses}

One of the main issues that were investigated in this work is the effect of
turning on masses for the adjoint fields. In the classical models (realized
in type IIA string theory) all possibilities were considered, with full
agreement between geometric (type IIA) and algebraic (field theoretical)
results. In the quantum models the classical branches of vacua may be lifted
by dynamically generated superpotentials. In the type IIA description this
is expected to correspond to forces between branes, as explained in subsection
\ref{force}.
Two kinds of situations were considered. Some comprehensive results were
obtained using M theory. In the following we will summarize the results and
discuss their implications both in field theory and in the dynamics of branes.

\subsubsection{\boldmath $N_L=N_R$, $\mu_L=-\mu_R$}

This situation is realized by a rotation of a detached $S_M$ (the central
$\NS$ brane) and was
investigated in subsection \ref{rot-flat}. Classically, there is a branch of
vacua parameterized by $\tilde{F}F$ moduli. One finds that this branch survives
also in the quantum model. In field theory, this follows from the conservation
of an $R$ symmetry, and in type IIA string theory this is because all $D_4$
branes extend between parallel $\NS$ branes ($S_L$ and $S_R$) and, therefore,
there are no forces between them. So we find agreement of all approaches.

Integrating out the massive adjoint fields, one obtains a model with a
Yukawa-like coupling between the fundamentals and the bi-fundamentals (see
eq. (\ref{Weff-m}))
\[ \frac{1}{\mu}\tr(\tilde{Q}BQ) \hsc
   B:=F\tilde{F}-\frac{1}{N}I_N\tr(F\tilde{F}) \hs, \]
so this procedure leads to the SW curves for this family of $N=1$ models.
For some of these models these curves were derived previously, using field
theoretical methods \cite{CEFS,Gremm9707}.
In all these cases, our results are in agreement with
the previous ones. For $n_Ln_R\neq0$ our results seem to be new.

Even when the results are known, the M theory approach often provides an
alternative understanding (as is already the case with the SW curve itself).
For example, the change of variables and the matching of scales necessary to
obtain a finite result in the limit of infinite mass follows naturally from
the fact that for a fixed scale, the ``rotation'' is actually a stretching.
Another example is the fact that the effective $U(1)$ gauge coupling is
independent of $\mu$ and tr$\tilde{F}F$. Geometrically this is obvious, since
these quantities describe the location and orientation of the detached $S_M$,
which has nothing to do with the gauge fields.

\subsubsection{\boldmath $\mu_L\neq0=\mu_R$} 

This situation is realized by the rotation of $S_L$ (the left $\NS$ brane) and
we analyzed in detail, in M theory (subsections \ref{rotate} and
\ref{non-deg}), the case with no fundamentals ($n_L=n_R=0$).
We found that the rotation is possible for any value of $m_F$ (the mass of the
bi-fundamentals), and it leads to a {\em complete lifting} of the $N=2$ Coulomb
branch, leaving only discrete vacua with no massless gauge fields.
Moving $S_R$ to the extreme right (which corresponds to vanishing coupling of
the $SU(N_R)$ gauge factor) we recovered (for infinite $\mu$)
the curve for $N=1$ SQCD (with a uniform quark mass). This agreement should be
regarded as a successful consistency check.

As remarked in subsection \ref{rotate}, the complete lifting of the Coulomb
branch appears in a quite general class of models: one
can consider a chain of an arbitrary number of $\NS$ branes connected by
$D_4$ branes, and what we found is that generically (\ie, when the curve is
irreducible) all parts of the moduli space are
lifted, including those related to remote
$D_4$ branes. This is a quantum effect:
returning, for concreteness, to the case of
two gauge factors, classically the
$A_R$ moduli remain free when one gives mass to $A_L$ (see subsection
\ref{mu-yn}), and this is supported
by the (classical) type IIA description, where the right $D_4$ branes are
suspended between parallel $\NS$ branes ($S_M$ and $S_R$) and, therefore, free
to move in the $v$-direction. 

\noindent{\bf Consequences for Field Theory:}

{}From the field-theoretical point of view
(considered in subsection \ref{FT-quant}), this
effect is expected to correspond, at scales below $\mu_L$, to the dynamical
generation of a
superpotential which depends on $A_R$ moduli and, in this way, constrains them.
In the limit of vanishing $\Lm_R$ and $\kp_R$ this superpotential should
reduce to that of $N=1$ SQCD \cite{IS9509} (see also \cite{HOO9706}).
But when either of these parameters is
non-vanishing, $A_R$ dependence is possible.
Unlike the previous case ($\mu_L=-\mu_R$), here the symmetries do not prevent
this possibility and we have shown in subsection \ref{FT-quant} that there is
a related model -- magnetic SQCD -- in which such a superpotential is
generated. The results from M theory mean that this is so also
in the present models.

There is another framework in which one can investigate the effect of a mass
for the adjoint: the low energy $N=2$ description for $\mu_L=0$.
It is interesting to see what our results mean in this framework.
One considers small
$\mu_L$ as a perturbation in the low energy $N=2$ description. Each point
in the moduli space has a neighborhood parameterized (holomorphically) by
moduli
$L_a,R_{\bar{a}}$, which are vev's of the scalar $N=2$ superpartners of the
massless gauge fields
(these are either the eigenvalues of $A_L$ and $A_R$, or related to them by
a duality transformation).  
The perturbation is $\mu_L U$, where $U$ is a function of
the moduli, that coincides with $\half\tr(A_L^2)$ in the classical limit.
At a generic point there are no other massless fields and the 
equations of motion require that the gradient of $U$ vanishes:
\[ \pt_aU=\pt_{\bar{a}}U=0 \]
which is (at least generically) false. To obtain a vacuum, one needs, for 
{\em every} non-vanishing component of the gradient, a massless $N=2$
hypermultiplet $(\tilde{q},q)$ charged {\em electrically}
under the corresponding gauge field
(this is what dictates the choice of moduli).
Then, the superpotential looks like
\[ W=\mu_L U+\sqrt{2}\tilde{q}Aq+\ldots \]
(where $A$ is the corresponding modulus), leading to the equations
\[ \mu_L \derp{A}U=-\tilde{q}q \hsc \tilde{q}A=Aq=0 \]
and implying $A=0$.

Naively, one would expect that $U$ is independent of
$R_{\bar{a}}$ (as is true classically),
leaving the right part of the Coulomb branch unlifted. But since these
parts are connected by the bi-fundamentals $\tilde{F},F$, this is not
necessarily true. Without such a restriction, generically all the
components of
the gradient do not vanish, implying that all the moduli are lifted.
We found that this is what happens, and this means that quantum effects ``mix''
the right and left moduli, leading to a dependence of $U$ on the $A_R$ moduli.
This is related to the left$\leftrightarrow$right mixing
in $P_L$, $P_R$ (see eq. (\ref{N=2-curve})),
which is possible in the parts proportional to the instanton factors
(recall that those parts were not determined by
the analysis in section \ref{M-th}).

\noindent{\bf Consequences for Brane Dynamics\footnote{We are grateful to 
David Kutasov for a discussion on this issue.}:}

The results from M theory have also implications on the dynamics of branes. 
Dynamically generated superpotentials in field theory
correspond to forces acting on the $D_4$ branes \cite{EGKRS9704}. This is
briefly described in subsection \ref{force}.
Such forces reduce the
possible equilibrium configurations of these branes and, therefore, lead to
the truncation of the moduli space. Once the equilibrium configurations are
known, they encode information about the forces that lead to them.

In the present case, in the equilibrium
configurations, as found in M theory (and translated to weakly coupled type
IIA language),
all the right $D_4$ branes are grouped together and the displacement
between them and the left $D_4$ branes (in the $v$ direction)
is a free parameter (corresponding to the mass $m_F$ of the
bi-fundamental in field theory).
The question is, what are the forces that lead to such equilibrium
configurations.
Following \cite{EGKRS9704}, there is a Coulomb-like 
attraction between the left and right $D_4$ branes. We should also impose the
constraint found in \cite{Witten9703}, that the average location of the $D_4$
branes is fixed. This constraint can be understood as a force between the 
$D_4$ branes and the $\NS$ branes on which they end. Combining these forces,
one obtains the correct vacuum structure (to show this, one uses the fact
that a Coulomb force is a {\em single valued} function of the
displacement). Note that without the constraint, the right $D_4$ branes
would all be 
concentrated at the same point (on the central $\NS$ brane) as the left
$D_4$ branes, which would mean that
the mass $m_F$ for the bi-fundamental is fixed
to vanish.
This is fully consistent with the corresponding field-theoretical description:
the constraint corresponds to the freezing of the $U(1)$ factor, so ignoring
it leads to a $U(N)$ gauge theory, where $m_F$ is a modulus (or,
more precisely, absorbed by a modulus $\tr A_R$) which is expected to
be determined by the (quantum) equations of motion (as are the other parts of
$A_R$). So we conclude that the constraint fixing the average position of the
$D_4$ branes should be imposed.

When there are more then three $\NS$ branes, we learn from the M theory
analysis that there are forces also on the rightmost $D_4$
branes, in spite of the fact that they are in direct
contact (\ie\ share a common $\NS$ brane) only with $D_4$ branes that extend
between parallel $\NS$ branes. We know that there are no forces between such
$D_4$ branes before the rotation. However, once a $\NS$ brane is rotated, 
there are forces acting on any $D_4$ brane connected to the rotated $\NS$ 
brane through other $\NS$ branes and $D_4$ branes (alternatively, there are 
``forces'' acting on any part of an $M_5$ brane connected to a ``rotated'' 
part; the ``non-locality'' in the action of forces in the type IIA
description is thus implied by the
holomorphicity of the fivebrane worldvolume in the other M theory limit).
When the chain is ``cut'' into
two disconnected components, only the $D_4$ branes
in the connected part of the rotated $\NS$ brane are affected by the rotation,
so this force is transfered through the brane configuration
(through the $M_5$ brane) and not through the bulk. 
This situation reminds a set of paramagnetic bodies put along a line. Normally,
there is no force between them, but when a magnet touches the body at the
left end of the line, this body is magnetized
and attracts the body to its right
and so on. When the magnet is taken away, the forces disappear.
Perhaps the forces
between $D_4$ branes that extend between parallel $\NS$ branes can be
formulated as such induced magnetization (or polarization).
The simplest scenario that leads to the correct vacuum structure would be
that the rotated $\NS$ brane induces (iteratively) attractive forces 
between $D_4$ branes extending between {\em the same} $\NS$ branes.
In a configuration with $N=2$ supersymmetry, the absence of these forces is
understood as a result of exact cancellations implied by the extended
supersymmetry. Such cancellations cannot be expected when this symmetry is
broken. It would be interesting to check this issue in other situations.

\subsection{More gauge group factors}

The considerations of this work extend naturally to models with more gauge
group factors (realized by brane configurations with more $\NS$ branes).
This was already discussed in the context of the lifting of the Coulomb
branch. The discussion of a degenerate $M_5$ brane can also be generalized
in this direction, leading to the determination of SW curves for more
$N=1$ SUSY models. For example, the detachment of a flat $\NS$
brane (one of the middle ones; in the absence of $D_6$ branes)
is possible whenever there are two adjacent factors with the same number of
colors. The rotation of the detached brane translates in field theory
to turning on masses for the corresponding two adjoint fields, and when these
fields are integrated out one obtains an $N=1$ model where not all the vector
superfields have a chiral $N=2$ superpartner.

\internote{Other points for discussion:
\nlb The Relevance of $x^6$
\nl identification of quark masses with the length of strings --- this length
changes when $D_6$ passes through a non-orthogonal $\NS$! discuss also the
limit of parallel $D_6$ and $\NS$.
\nlb Geometric Realization of the Yukawa Couplings;
$\kp,\lm$ dependence on $\th$:
\nl Some observations appear in section \ref{Yukawa}
\nl independence known in FT
\nl additional information from section 4.4.2 (see also sec B.4.3)
\nl did we assume $\kp,\lm\neq  0,\infty$?
\nlb Correct asymptotic conditions
\nl $m_F$ not determined;
\nl infinite branes that can be finite (letter 3/8).}

\vspace{1cm}
\noindent{\bf Acknowledgment:}
We are grateful to Shmuel Elitzur, David Kutasov,
Yaron Oz and Eliezer Rabinovici for
helpful discussions.
This work is supported in part by BSF -- American-Israel Bi-National
Science Foundation, and by the Israel Science Foundation founded by the
Israel Academy of Sciences and Humanities -- Centers of Excellence Program.
A.G. thanks the Theory Division at CERN for hospitality.

\vspace{1cm}
\noindent{\bf Note Added:}

After completion of this work, we received the preprint \cite{LPT9708} which
also considers the Coulomb branch of some $N=1$ SUSY gauge models with product 
groups by type IIA brane configurations.


\appendix
\renewcommand{\newsection}[1]{
 \vspace{10mm} \pagebreak[3]
 \refstepcounter{section}
 \setcounter{equation}{0}
 \message{(Appendix \thesection. #1)}
 \addcontentsline{toc}{section}{
  Appendix \protect\numberline{\Alph{section}}{#1}}
 \begin{flushleft}
  {\large\bf Appendix \thesection. \hspace{5mm} #1}
 \end{flushleft}
 \nopagebreak}


\newsection{Symmetries}
\secl{sym}

The classical
$SU(N_L)\otimes SU(N_R)$ model with vanishing superpotential has a large
group of global symmetries, which include non-Abelian factors 
$SU(n_L)^2\otimes SU(n_R)^2$, and $U(1)$ factors described in the following
table%
\footnote{For $N_L=N_R$ and $n_L=n_R$ there is also a $\ZZ_2$ symmetry
  exchanging $L\leftrightarrow R$.}
\internote{In the $U(N_L)\otimes U(N_R)$ model, there are two additional $U(1)$
  factors transforming the singlet parts of $A$. These will not preserve the
  form we chose for the superpotential and seem not to be useful for the
  present work.}
\beq\begin{array}{|c||c|c|c|c|c|c|c|c|c||c|c|}
  \hline \rule[-1.3ex]{0em}{4ex}
        &\th&F&\tilde{F}&A_L&A_R&Q_L&\tilde{Q}_L&Q_R&\tilde{Q}_R&\Lml  &\Lmr \\
  \hline
  q_\th & 1 & &         &   &   &   &           & &  &-2(n_L+N_R)&-2(n_R+N_L)\\
  q_F   &   &1&       1 &   &   &   &           & &  &      2N_R &      2N_L \\
  q_{FV}&   &1&      -1 &   &   &   &           & &  &           &           \\
  q_{AL}&   & &         & 1 &   &   &           & &  &      2N_L &           \\
  q_{AR}&   & &         &   & 1 &   &           & &  &           &      2N_R \\
  q_L   &   & &         &   &   & 1 &         1 & &  &      2n_L &           \\
  q_{LV}&   & &         &   &   & 1 &        -1 & &  &           &           \\
  q_R   &   & &         &   &   &   &           &1& 1&           &      2n_R \\
  q_{RV}&   & &         &   &   &   &           &1&-1&           &           \\
  \hline\end{array}
\eeq 
where $\th$ is the Fermionic coordinate of superspace and
$\Lml,\Lmr$ are the instanton factors of the two gauge groups.
The charges of the instanton factors $\Lml,\Lmr$ are chosen to 
  compensate the anomalies. To determine them, one needs the Dynkin indices
  $k$ of the $SU(N)$ gauge groups: $k=N$ for the adjoint and $k=\half$ for the
  fundamental. Similarly, the charges of the parameters in the action can be
  determined by the requirement that the (classical) action is invariant
  (recall that this means that the superpotential has R-charge $q_\th=2$).

These symmetries can be used to constrain the possible form of
superpotentials and SW curves. Here we describe our strategy of using them. 
\begin{itemize}
\item Conservation of $q_{FV}$, $q_{LV}$ and $q_{RV}$ (the ``baryon numbers'')
means that the corresponding fields must appear in pairs
$\tilde{F}^{\bar{a}}_{\hsm a}F^b_{\hsm\bar{b}}$,
$\tilde{Q}^{\hsm i}_{L\hsm a}{Q_L}^b_{\hsm j}$, 
$\tilde{Q}^{\hsm\bar{i}}_{R\hsm\bar{a}}{Q_R}^{\bar{b}}_{\hsm\bar{j}}$.

\item In most parts of this work we set the Yukawa couplings $\kp$ and $\lm$
to fixed values (either 1 or $\sqrt{2}$). The conservation of
$q_{AL},q_{AR},q_L+q_{LV}$ and $q_R+q_{RV}$ can be used to recover
$\kp_L,\kp_R,\lm_L$ and $\lm_R$ respectively.
In fact, an efficient way to insure the conservation of the above charges is
to use only neutral quantities. The prescription described in eqs.
(\ref{kp},\ref{lm}), can be interpreted as the identification of a neutral
expression that reduces to the correct quantity when $\kp$ and $\lm$ are set
to 1. For example $\mu_L$ represents the ($q_{AL}$-neutral) expression
$\mu_L/\kp_L^2$. In the quantum theory additional quantities
appear -- the instanton factors and the coordinates of the SW curve. With the
above approach, the coordinates are chosen to be neutral and the instanton
factors (for $\kp=\lm=1$) represent
\beql{kplm} \Lm^b\goto\kp^{2N-n}\Lm^b\det\lm \hs. \eeq
When the Yukawa couplings are fixed to a value different from 1, the
transformations recovering them should be modified in an obvious way. See, for
example, eq. ({\ref{recover}}).

\item We are left with two unused symmetries. To be useful when $\kp$ and 
$\lm$ are fixed, these symmetries should be chosen not to act on $\kp$ and
$\lm$. One such choice is $R_{45}$ and $R_{89}$, defined in table (\ref{U1RT}).

\end{itemize}

\ifdraft\beginsup
\input{sunsup}
\endsup\fi


\end{document}